\documentclass[aip,cha,preprint,numerical]{revtex4-1}
\usepackage{subfig}
\usepackage{graphicx}
\usepackage{amsmath}

\draft 

\begin{document}


\title{Generating self-organizing collective behavior using separation dynamics from experimental data} 



\author{Graciano Dieck Kattas}
\affiliation{Department of Electronic and Information Engineering \\
The Hong Kong Polytechnic University, Hung Hom, Kowloon, Hong Kong}
\author{Xiao-Ke Xu}
\affiliation{Department of Electronic and Information Engineering \\
The Hong Kong Polytechnic University, Hung Hom, Kowloon, Hong Kong}
\affiliation{School of Communication and Electronic Engineering \\
Qingdao Technological University, Qingdao 266520, China}
\author{Michael Small}
\email[]{michael.small@uwa.edu.au}
\affiliation{School of Mathematics and Statistics, University of Western Australia, Crawley, WA 6009, Australia}
\affiliation{Department of Electronic and Information Engineering \\
The Hong Kong Polytechnic University, Hung Hom, Kowloon, Hong Kong}


\date{\today}

\begin{abstract}
Mathematical models for systems of interacting agents using simple local rules have been proposed and shown to exhibit emergent swarming behavior. Most of these models are constructed by intuition or manual observations of real phenomena, and later tuned or verified to simulate desired dynamics. In contrast to this approach, we propose using a model that attempts to follow an averaged rule of the essential distance-dependent collective behavior of real pigeon flocks, which was abstracted from experimental data. By using a simple model to follow the behavioral tendencies of real data, we show that our model can exhibit emergent self-organizing dynamics such as flocking, pattern formation, and counter-rotating vortices. The range of behaviors observed in our simulations are richer than the standard models of collective dynamics, and should thereby give potential for new models of complex behavior.
\end{abstract}


\maketitle 


\begin{quotation}
We propose a simple dynamical model of collective behavior that attempts to follow a rule of local neighbor interactions, which was abstracted from experimental flight data of pigeon flocks. This rule is consistent with the previous hypothesis of the basic mechanisms affecting collective motion: short range repulsion to avoid collisions, longer range attraction to keep the group together, and velocity alignment to maintain the same navigational direction. The local interactions of our model consist of using naive neighbor estimates: essentially assuming that nearest neighbors will move with the same velocity as in the previous time interval. The dynamics of our model try to follow the experimental rule by using a simple adjustment mechanism with respect to the naive neighbor estimates. From our simulations we show that by changing the initial conditions or the number of individuals involved in the interactions, the model is capable of exhibiting a wide range of realistic behaviors, significantly richer than the dynamics observed in traditional models. Our study emphasizes the importance of using experimental data for making better models of complex systems, and that this should contribute to a better understanding of nonlinear collective dynamics.       

\end{quotation}

\section{Introduction}

Multi-agent dynamical systems with collective behavior have attracted many recent studies, especially arising from the scientific interest in animal movement and interactions. Commonly called ``swarming", the aggregation and coordination of animals moving together is a captivating phenomenon that exhibits intelligent behavior and evolutionary properties, which benefit the group as a whole. Such cohesive and synchronized group motion, is essential to guide a population through a possibly dangerous environment with many irregularities that can definitely affect individual movement. This type of collective behavior has been observed and studied in birds, insects, fish, and mammal herds \cite{Okubo2001, Potts1984}. Research has led to differences regarding the interactions between individuals, in particular, the navigational force of a swarm can be dominated by informed ``leaders" \cite{Couzin2005}, or it could simply be a consequence of democratic local interaction rules between neighbors \cite{Conradt2003}. The latter is sometimes referred as the ``many wrongs" principle, and it asserts that the navigational errors in a group are removed by tight cohesion and mutual interaction between individuals \cite{Simons2004, Codling2007}. For the particular case of bird flocking, recent advances in measurement technology have allowed convincing conclusions to be obtained regarding the existence of leaders \cite{Nagy2010}, as well as the topological nature of the local interactions between neighboring birds \cite{Ballerini2008}.

Many types of mathematical models have been proposed to simulate collective dynamics. Attraction, short-range repulsion to avoid collisions, and velocity alignment, have been typically characterized as the main mechanisms to describe swarming behavior in models of animal groups \cite{Okubo2001}. These rules, influenced by the local interactions between individuals are the same principle as that behind the well-known ``Boids" model \cite{Reynolds1987}, which has been traditionally used to generate swarm animations. The Vicsek model showed that simple orientation alignment between local neighbors is enough to generate complex collective behavior \cite{Vicsek1995}. Newer modeling approaches have considered different attraction-repulsion-alignment mechanisms and generated a variety of more complex collective dynamics, such as individuals rotating around a center like a vortex (also called milling) \cite{Couzin2002,Erdmann2005,Chuang2007,Touma2010}.
Others models have used a set of ``informed" individuals or leaders with superior information in order to drive the direction of the swarm \cite{Couzin2005, Raghib2010,Zhang2010}. All these models have been based on the principle of manually tuning their parameters to get a desired collective behavior in the dynamics, or to compare between a wide variety of tendencies in the simulations. New modeling approaches have considered positional data to automatically tune parameters in a fixed model structure, in order to have behavior based on experimental observations \cite{Lukeman2010, Eriksson2010}. The approach we advocate here is to begin with actual field measurements of animal collective behavior and to ask what rules one can deduce directly from that data. The recent availability of extensive experimental data motivates this direct analysis, in order to infer the true mechanisms of swarming in nature.         

We adhere to the philosophy of using data to construct models capable of simulating collective dynamics \cite{Kattas2011b}, inspired by approaches that have been proposed before in other areas and contexts to infer dynamical systems \cite{Judd1995,Small2002,Bongard2007,Gennemark2007}, or their natural laws \cite{Schmidt2009}. Using a dataset of positional data from GPS \cite{Nagy2010}, in a previous study we built models that showed the importance of mutual local interactions in pigeon flocks \cite{Kattas2011b}, which is consistent with the ``many wrongs" principle. From the same study, using our models we inferred an averaged local interaction rule that is analogous to the attraction, short-range repulsion, and orientation alignment mechanisms from earlier models. Similar behaviors have been inferred for shoaling fish recently in another work \cite{Herbert-Read2011}. In this paper, we show that by using a new abstracted model in which the individuals of the swarm use a simple adjustment mechanism to follow this natural rule, it is possible to generate complex and realistic self-organizing dynamics. While many of the individual behaviors our model produces have been reported before, such as mobile cohesive motion \cite{Couzin2002}, counter-rotating vortices \cite{Erdmann2005,Chuang2007,Touma2010}, initially aligned individuals losing their unified movement to uncertain oscillations \cite{Erdmann2005}, and pattern formation \cite{Cheng2011}, to our knowledge, it is the first one able to generate dynamics resemblant to all of them by simply varying the density and the number of interacting neighbors in the simulations. This new variety of behaviors emphasizes the emergence of complex phenomena from simple collective interactions and it should open the door to a new generation of models with dynamics from behavioral rules extracted from actual experimental data. 

\section{Separation dynamics from experimental data}

In our previous study \cite{Kattas2011b}, using a general flocking model built from data of several homing pigeon flights \cite{Nagy2010}, we measured collective behavior by calculating averaged attraction/repulsion statistics that quantify the change in separation of nearest neighbors after a time interval. We shall denote this as ``separation dynamics", and describe it by showing how the averaged separation between neighbors at a given time, $\xi(t)$, affects the change at the next interval $\Delta \xi(t)$. We considered a fixed number of nearest neighbors for interactions, due to a recent contribution which showed that birds interact with a specific number of neighbors rather than with all within a radius \cite{Ballerini2008}. From simulations of our models built from experimental data, we averaged $\xi(t)$ and  $\Delta \xi(t)$ values for different density and speed scenarios \cite{Kattas2011b}. Our analysis produced a curve that showed repulsion tendencies ending at shorter separations than 20 m, followed by attraction between 20 and 500 m, reaching a maximum strength near 120 m. Figure 1(a) shows a cubic spline interpolation of the retrieved curve, which will be used in this paper as the main rule for the new simplified model. In addition, Figure 1(b) shows speed statistics (magnitude of velocity) also averaged for the same separation values. The described behaviors are consistent with the basic demeanors of collective dynamics in classical models: repulsion at short separations to avoid collisions, attraction at longer separations to keep group cohesion, and velocity alignment to have a synchronized swarm moving together. The latter behavior can be confirmed by visualizing in the plots that near 20 m of separation, we have weak near-zero attraction/repulsion, but a high speed; implying steady separations but fast movement, thus equal orientations in the velocities of the interacting individuals. We must emphasize that the exact shapes of the curves are not important for the new model to be described in the next section, as long as the same described qualitative properties (short-range repulsion, attraction, speed) and continuous forms are roughly similar.  

\begin{figure}[htb]
  \centering
  \subfloat[Attraction(-)/repulsion(+)]{\label{fig:figure1a}\includegraphics[scale=0.52]{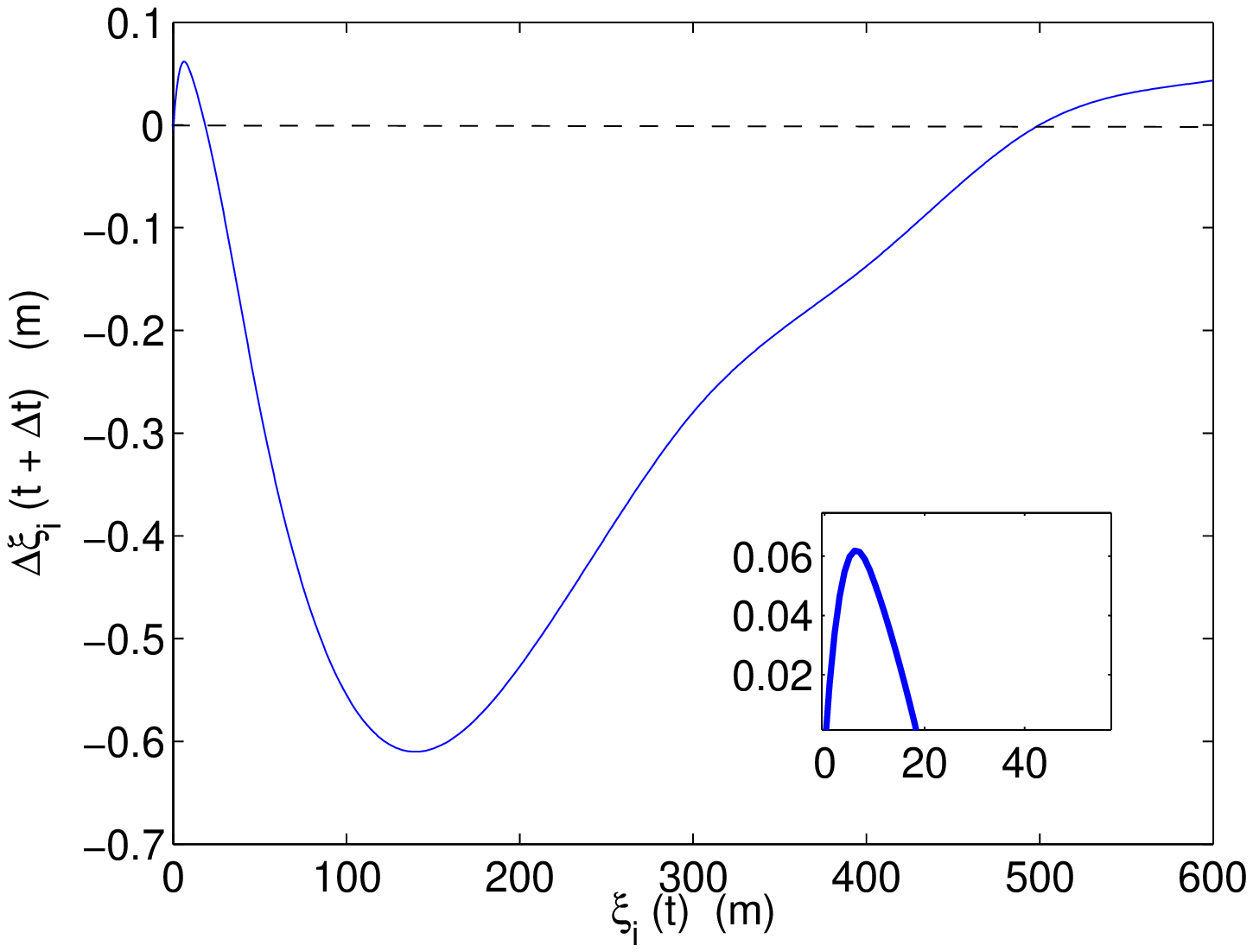}}             
  \hspace{0.1cm}   
    \subfloat[Speed]{\label{fig:figure1b}\includegraphics[scale=0.52]{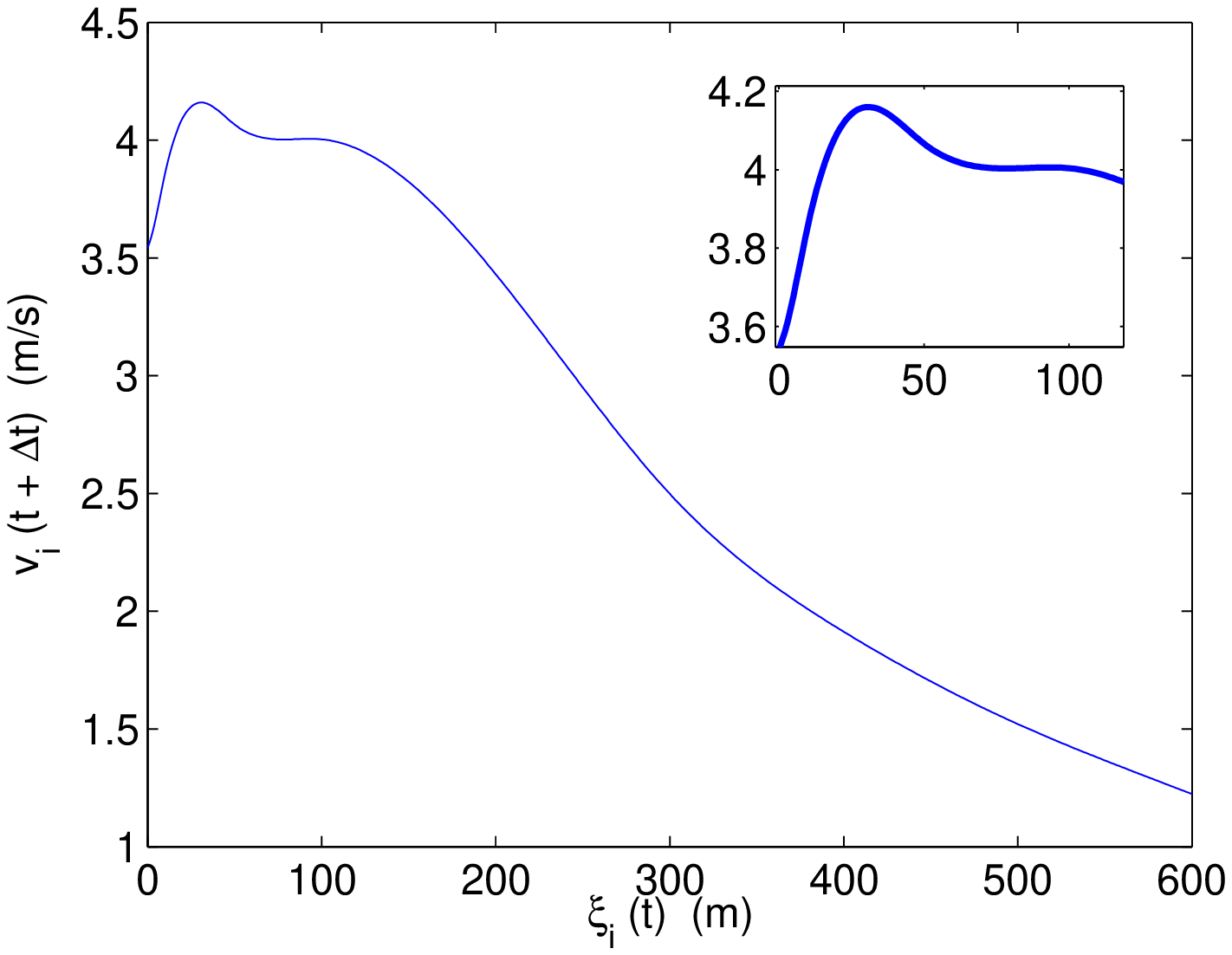} }
  \caption{Attraction/repulsion and speed curves obtained from cubic spline interpolations of our retrieved statistics \cite{Kattas2011b}. The plots show change in separation (a) and speed (b) for the next time interval ($t+\Delta t$) as a function of average separation to neighbors of an individual \emph{i} at time \emph{t}, denoted $\xi(t)$. Sample time $\Delta t$ is 2 seconds. }
  \label{fig:figure1}
\end{figure}

\section{Naive model using separation dynamics} 

Using the separation dynamics presented in the previous section, we propose to build an abstracted 2D model which follows the curves to an extent, especially the attraction/repulsion tendencies. Our motivation is to keep the model as simple and tractable as possible, in order to present it as a basic reference of how complex behavior can arise from relatively simple models that follow tendencies extracted from real experimental data. First of all, we shall define the attraction curve as function $f[\xi(t)]$ and the speed function as $v[\xi(t)]$; in addition, we denote \emph{N} individuals, \emph{M} nearest neighbors, and an infinite space for movement (no boundary conditions). When considering the update for an individual \emph{i}, trying to follow the attraction rule with respect to \emph{M} neighbors can be quite difficult to compute, and has many spatial possibilities. To simplify matters, but still keeping faithful to the original concept, we treat the \emph{M} neighbors interacting with an individual \emph{i} as a single averaged entity, and thus we shall use the separation of individual \emph{i} to the average position of its neighbors at a time interval, $\delta_i(t)=\| \textbf{x}_i(t)-\langle \textbf{x}_j(t) \rangle_{M} \|$, as the separation measure to be considered for both curves, and thus we take $\delta_i(t)$ as $\xi_i(t)$. We note that even though the distance between an individual and the center of mass of its nearest neighbors ($\delta_i(t)$) is not equal to the the average distance between an individual and its nearest neighbors ($\xi_i(t)$), it is a meaningful distance to most of the other neighbors when considering irregularly distributed groups, and thus a reasonable simplification.

The position update for an individual \emph{i} in the model proceeds in two steps, the first one consisting of a preliminary update based on just the speed curve: 
\begin{equation}
\textbf{x}_i(t+\Delta t)'= \textbf{x}_i(t) + {v[\delta_i(t)]}{\textbf{u}_i(t)}{\Delta t}
\end{equation}
where $\textbf{u}_i(t)$ is the unit vector of the previous update, $\Delta \textbf{x}_i(t)=\textbf{x}_i(t)- \textbf{x}_i(t-\Delta t) $, and therefore giving $\textbf{u}_i(t)=\frac{\Delta \textbf{x}_i(t)}{\| \Delta \textbf{x}_i(t) \|}$. With this first step, we are calculating a preliminary update which is simply based on keeping the previous orientation (the unit vector), but using the new speed value which corresponds to the separation to neighbors at time \emph{t}, that is $v[\delta_i(t)]$. 

The second step of our position update  consists of ``adjusting" the preliminary update in accordance to the attraction curve. Since we treat the \emph{M} neighbors interacting with \emph{i} as a single averaged entity, it can be defined as $\textbf{y}_i(t)=\langle \textbf{x}_j(t) \rangle_M $. Taking this into account, we define the naive neighbor estimate as: 
\begin{equation}
\textbf{y}_i(t+\Delta t)'=\textbf{y}_i(t) + \Delta \textbf{y}_i(t)
\end{equation}
where $\Delta \textbf{y}_i(t)=\textbf{y}_i(t)- \textbf{y}_i(t-\Delta t) $, which basically is an assumption that all neighbors will move with the same velocity as in the previous iteration. We call it ``naive" in allusion to the traditional ``naive predictor" in time series analysis, that consists in using the current value as a forecast of the next one.  Afterward, we estimate the next separation directly from the attraction curve, that is:
\begin{equation}
 \delta_i(t+\Delta t)'= \delta_i(t) + f[\delta_i(t)]
\end{equation}
Then we calculate the ``adjustment", which is a scalar value representing half the distance required to reach the estimated next separation, between the preliminary update and the naive neighbor estimate:
\begin{equation}
a_i(t)= \frac{1}{2}  [\| \textbf{y}_i(t+\Delta t)' - \textbf{x}_i(t+\Delta t)' \| -\delta_i(t+\Delta t)' ] 
\end{equation}
Finally, we ``adjust" the preliminary update and add noise:
\begin{equation}
\textbf{x}_i(t+\Delta t)= \textbf{x}_i(t+\Delta t)'  +  {a_i(t)} {\textbf{m}_i(t)} + {\boldsymbol{\eta}(t)} 
\end{equation}
where $\textbf{m}_i(t)$ is a unit vector pointing from the preliminary update to the naive neighbor estimate, $\textbf{m}_i(t)=\frac{ \textbf{y}_i(t+\Delta t)' - \textbf{x}_i(t+\Delta t)' }{\| \textbf{y}_i(t+\Delta t)' - \textbf{x}_i(t+\Delta t)' \|} $; and $\boldsymbol{\eta}_i(t)$ is a random vector that represents the noise in the update.

In simple words, the ``adjustment" consists of considering all neighbors as a single averaged entity, and fixing the separation between the preliminary update $\textbf{x}_i(t+\Delta t)' $  and the naive neighbor estimate $\textbf{y}_i(t+\Delta t)'$, to attempt to follow the new separation obtained from the attraction curve $\delta_i(t+\Delta t)'$. Of note here is that the $\frac{1}{2}$ in adjustment $a_i(t)$ denotes half of the required magnitude to reach the desired separation, because we loosely assume that the averaged neighbor entity will contribute with the other half. The idea behind the simplification is to preserve the attraction rule as if individual \emph{i} is interacting with a single neighbor (which is actually the center of mass of the neighbors). Under steady state conditions, which means $M+1$ individuals moving with the same velocity and optimal separations ($f[\delta_i(t)]=0$), we can expect adjustment $a_i(t)$ to be zero, since the naive neighbor estimate will be accurate, and thus the separation between the preliminary update and the naive neighbor estimate will be roughly equal to the desired separation: $ \| \textbf{y}_i(t+\Delta t)' - \textbf{x}_i(t+\Delta t)' \| \approx  \delta_i(t+\Delta t)'$.

\section{Results}

\subsection{Description of system behaviors}

Extensive simulations using different sets of parameter values were carried out to explore the dynamics of this model. We verified that for a fixed number of individuals, the key parameters that influence model behavior are the initial positional density (analogous to the Visek paper \cite{Vicsek1995}), the number of nearest neighbors considered in the interactions (\emph{M}), and the orientations of the initial velocities. For all our simulations, the initial density was varied by modifying a parameter $r_c$, which symbolizes the radius of the circle where the initial positions were randomly distributed (uniformly). The noise of the update was varied according to parameter $\epsilon$ in $\boldsymbol{\eta}_i(t)=\textbf{U}(-\epsilon,\epsilon)$. The initial orientations were either set randomly or equal (aligned) for all individuals, to compare the differences between initially directed and undirected states in the population.

First of all, some of the most significant behaviors that were observed to be repeatable under the same parameter settings, will be described. Afterwards, a general outlook of the system behavior for the different parameter values will be outlined as a final picture to describe phase transitions and capabilities of the model. All the descriptions and plots are restricted to simulations of $N=100$ individuals, and with $\Delta t = 2$, the latter in accordance to the experimental attraction curve. The noise was set to $\epsilon=0.5$ because it was found to be a low enough value that does not dominate the deterministic dynamics of the system, but still able to give enough variation for the system to show robust pattern formation and self-organization, e.g. individuals ``moving around" within a shape. We note that the parameter values that we use for the simulations described in this section are not restrictions or strictly necessary to reproduce these same behaviors. We chose a parameter space that based on our observations of the simulations, showed an interesting range of behaviors and conditions that were worthy to analyze. 

\begin{figure}[htb]
  \centering
  \subfloat[Local swarming]{\label{fig:figure2a}\includegraphics[scale=0.52]{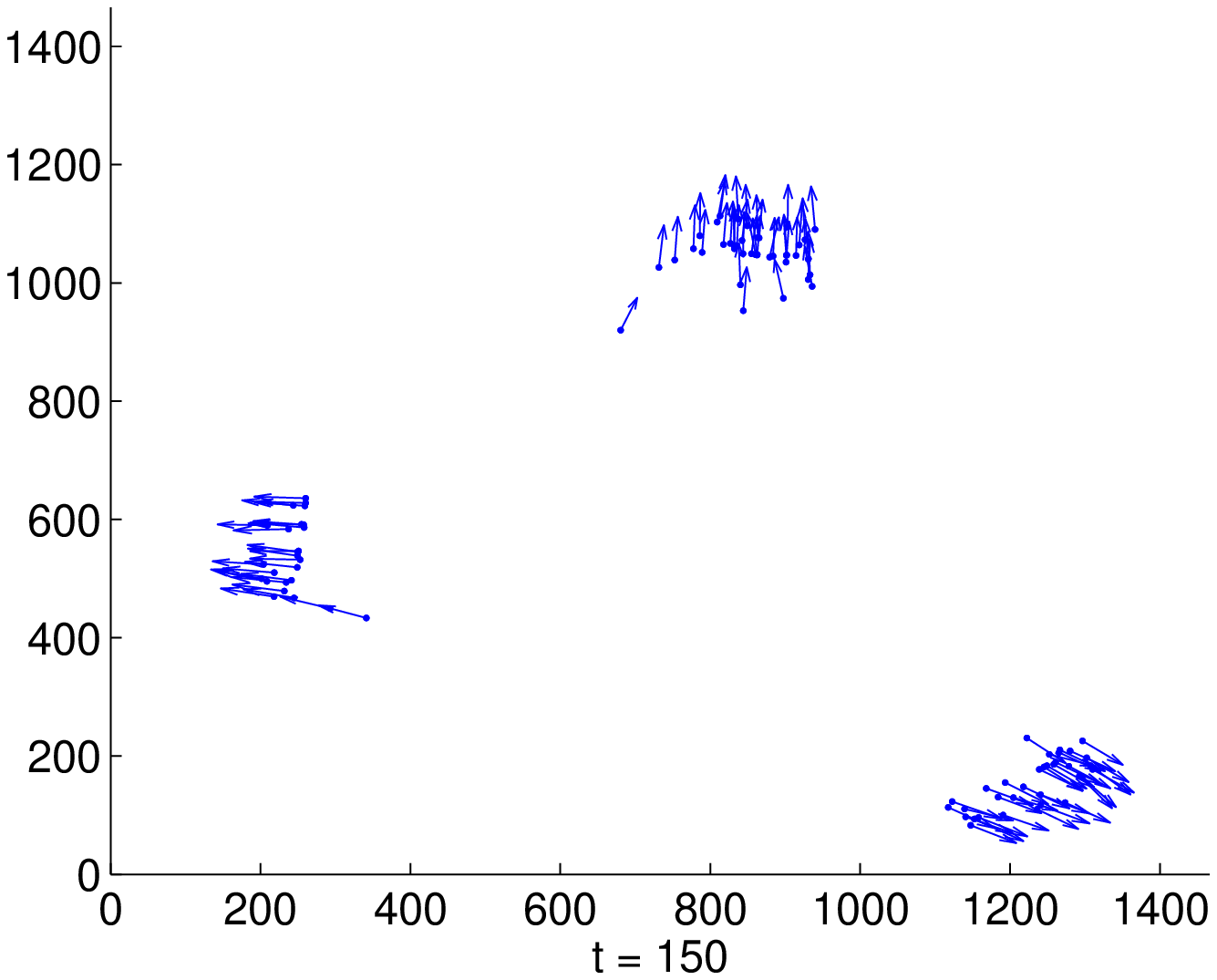}}             
  \hspace{0.1cm}   
  \subfloat[Unstable triangle]{\label{fig:figure2b}\includegraphics[scale=0.52]{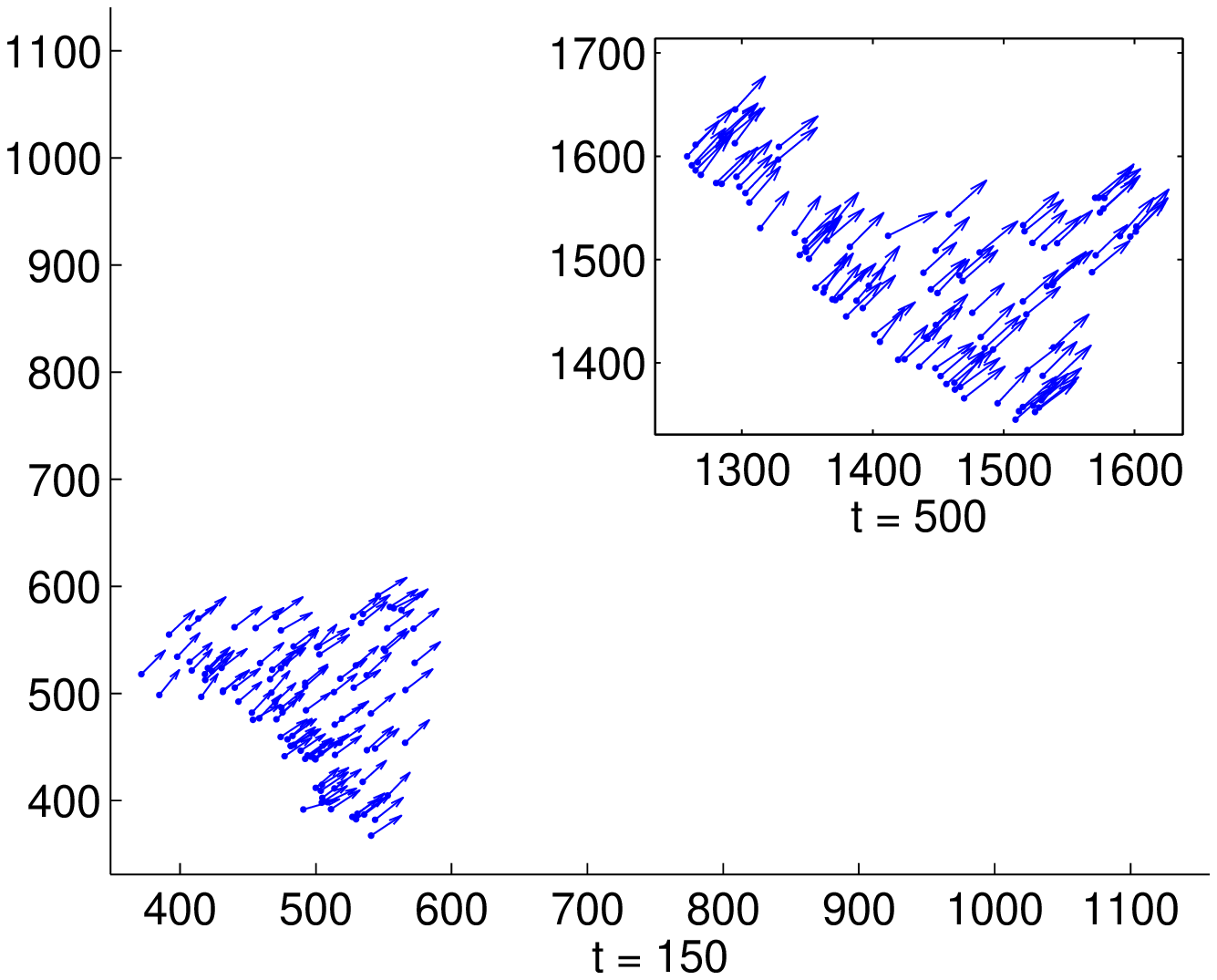} }
  \caption{Simulations with $M=15$, $r_c=100$, $\epsilon=0.5$. Random initial orientations in (a), snapshot at $t=150$. Initially aligned orientations in (b), snapshots at $t=150$ and $t=500$  (enhanced online: sim2a.mp4 , sim2b.mp4)}
  \label{fig:figure2}
\end{figure}

For small values of \emph{M} (between 10 and 20), $r_c$ between 50 and 400, and random initial orientations,  stable local swarming occurs in small groups, which is expected considering the small number of interacting neighbors. Figure 2(a) shows a simulation with $M=15$ where the population is split into three stable groups that move away. For similar parameter values but aligned initial orientations, we consistently observed an interesting behavior where the whole population initially self-organizes into a loose triangle, which turns out to be unstable and disintegrates as the simulation advances. Figure 2(b) shows an example of how such a split can occur.   

Global cohesive movement of the whole population was obtained for values of $M \geq 30$, but this depended largely on the initial density and orientations. Essentially mid-range high density conditions with $r_c$ values near 200, give the best cohesion. For $M=30$ and values of $r_c$ between 200 and 400, and random initial orientations, the group moves together in a single direction after a transient semi-synchronous state, and self-organizes into a stable shape that is very resistant to the noise. An example can be seen in Figure 3(a). Using initially aligned orientations, high \emph{M}, and no noise, the population constantly self-organizes into more regular shapes with interesting features, e.g. the roughly rectangular, circular, and linear sections in Figure 3(b).  

\begin{figure}[htb]
  \centering
  \subfloat[Stable cohesive unit]{\label{fig:figure3a}\includegraphics[scale=0.52]{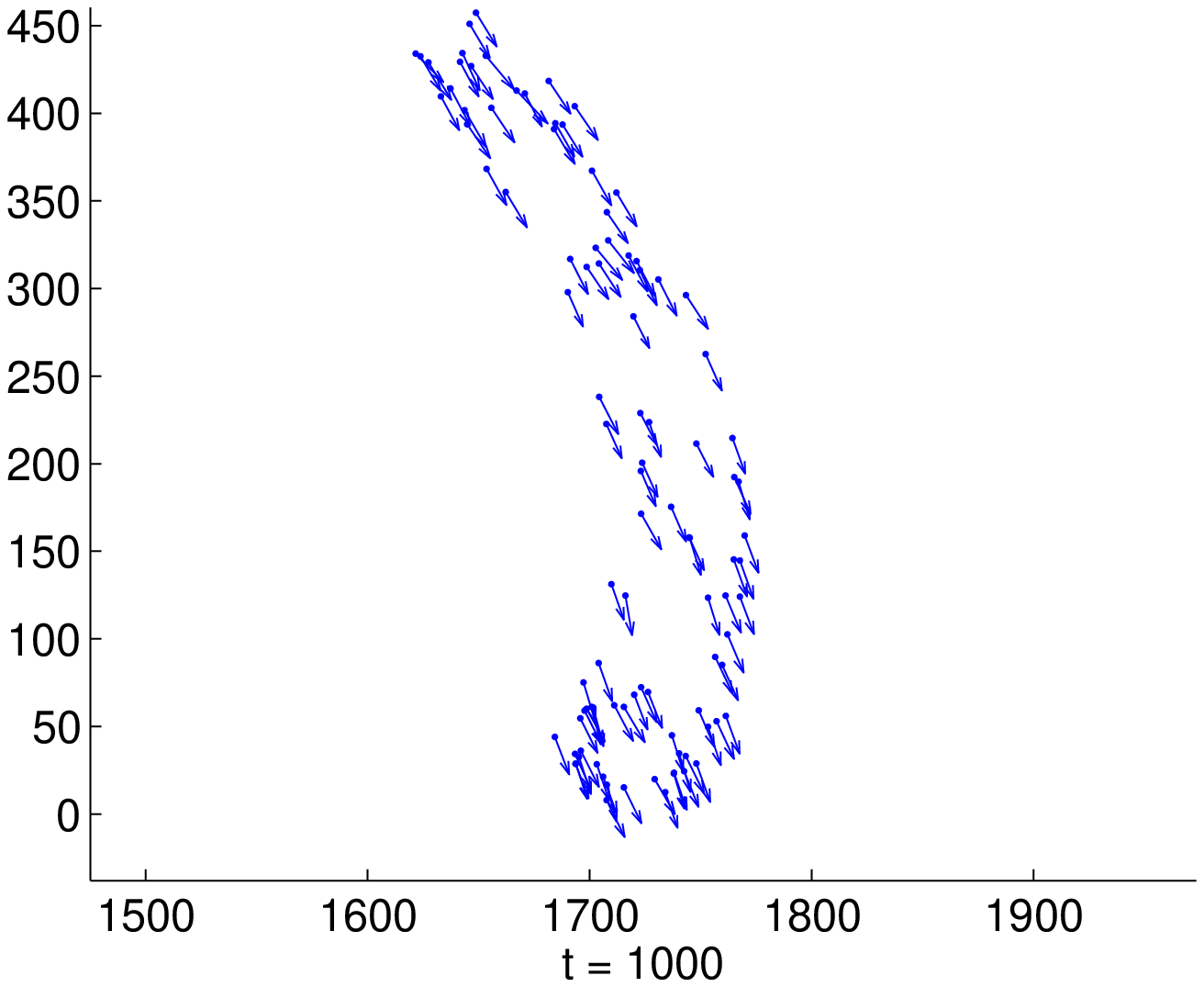}}             
  \hspace{0.1cm}   
  \subfloat[Regularized self organization]{\label{fig:figure3b}\includegraphics[scale=0.52]{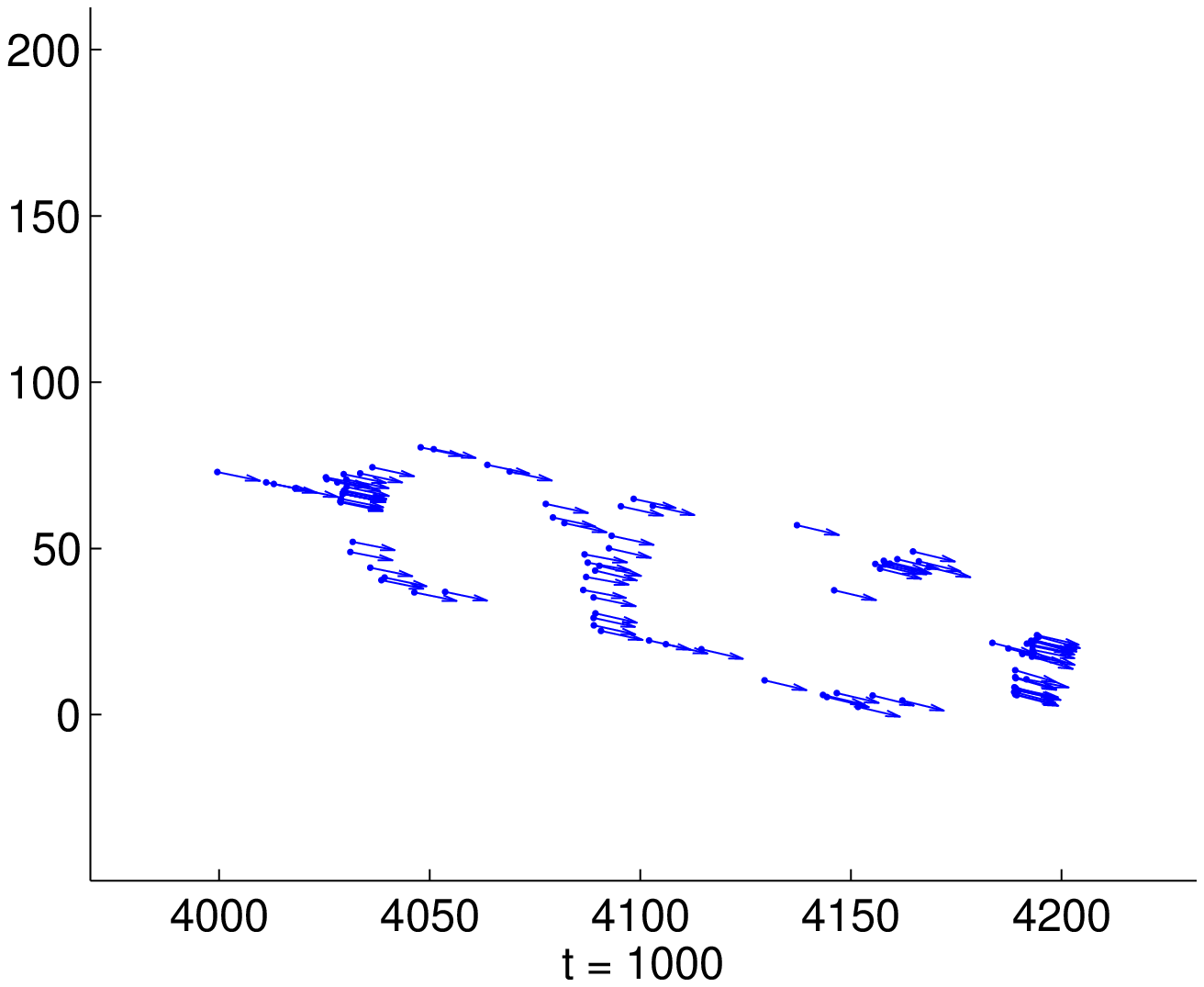} }
  \caption{Simulations with $r_c=200$. $M=30$, $\epsilon=0.5$ and random initial orientations in (a), snapshot at $t=1000$. $M=50$, $\epsilon=0$ and initially aligned orientations in (b), snapshot at $t=1000$ (enhanced online: sim3a.mp4, sim3b.mp4) }
  \label{fig:figure3}
\end{figure}

A very interesting complex behavior emerges by decreasing the initial density (increasing $r_c$), and maintaining a high \emph{M}. Dynamics with particles counter-rotating around a moving center, resembling a vortex, was observed for $r_c \geq 400$, $M=50$ and random initial orientations. Figure 4(a) shows a plot of a snapshot during a simulation. The dynamics involve individuals further away from the center slowly trying to approach the center with rotations. A portion of the population rotates clock-wise around an empty center, while the rest does the same but counterclockwise. The empty center of the group, with individuals rotating around it, slowly moves in random directions that depend on the overall flow, resembling the behavior of many natural collective phenomena.

\begin{figure}[htb]
  \centering
  \subfloat[Vortex]{\label{fig:figure4a}\includegraphics[scale=0.52]{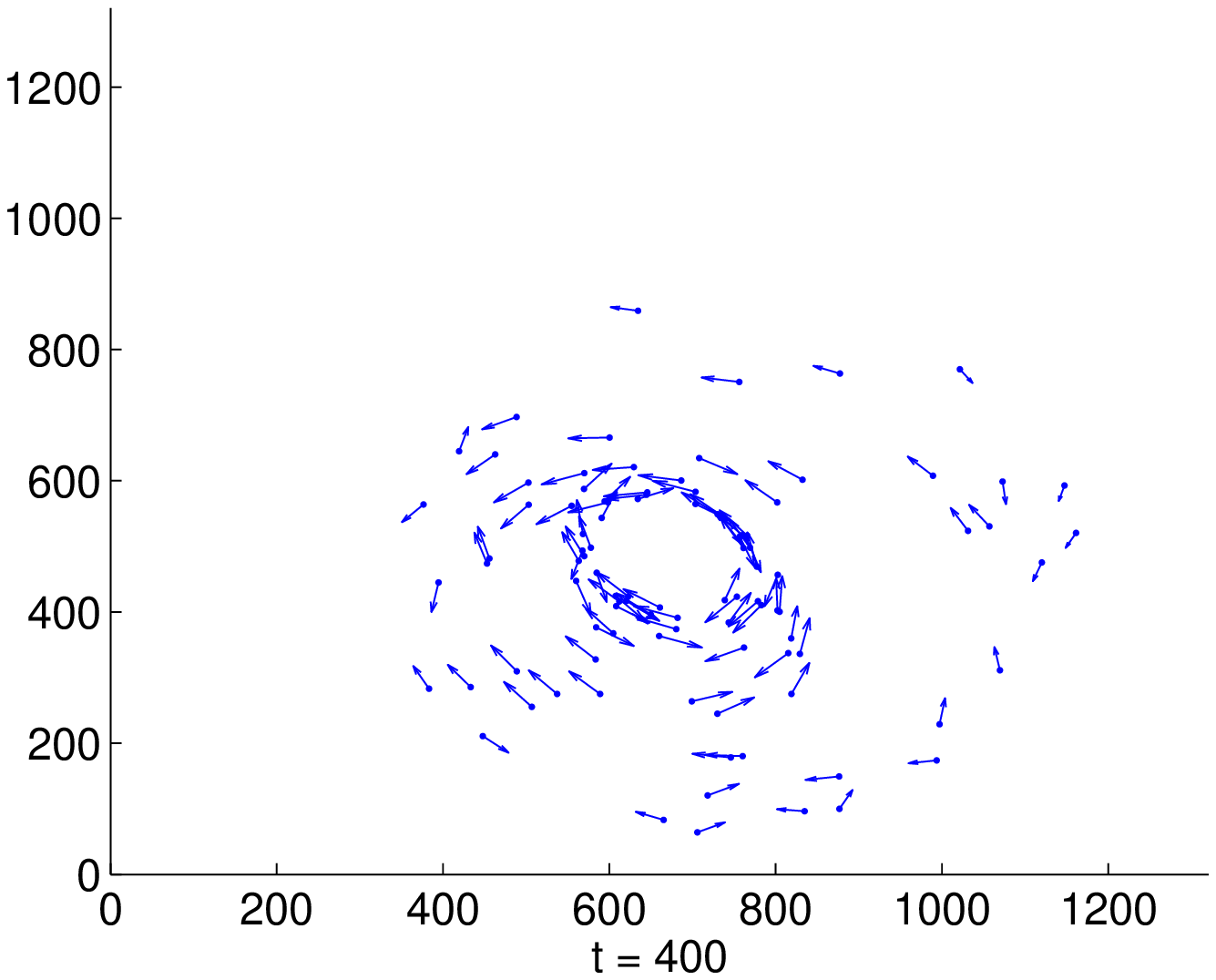}}             
  \hspace{0.1cm}   
  \subfloat[Comet]{\label{fig:figure4b}\includegraphics[scale=0.52]{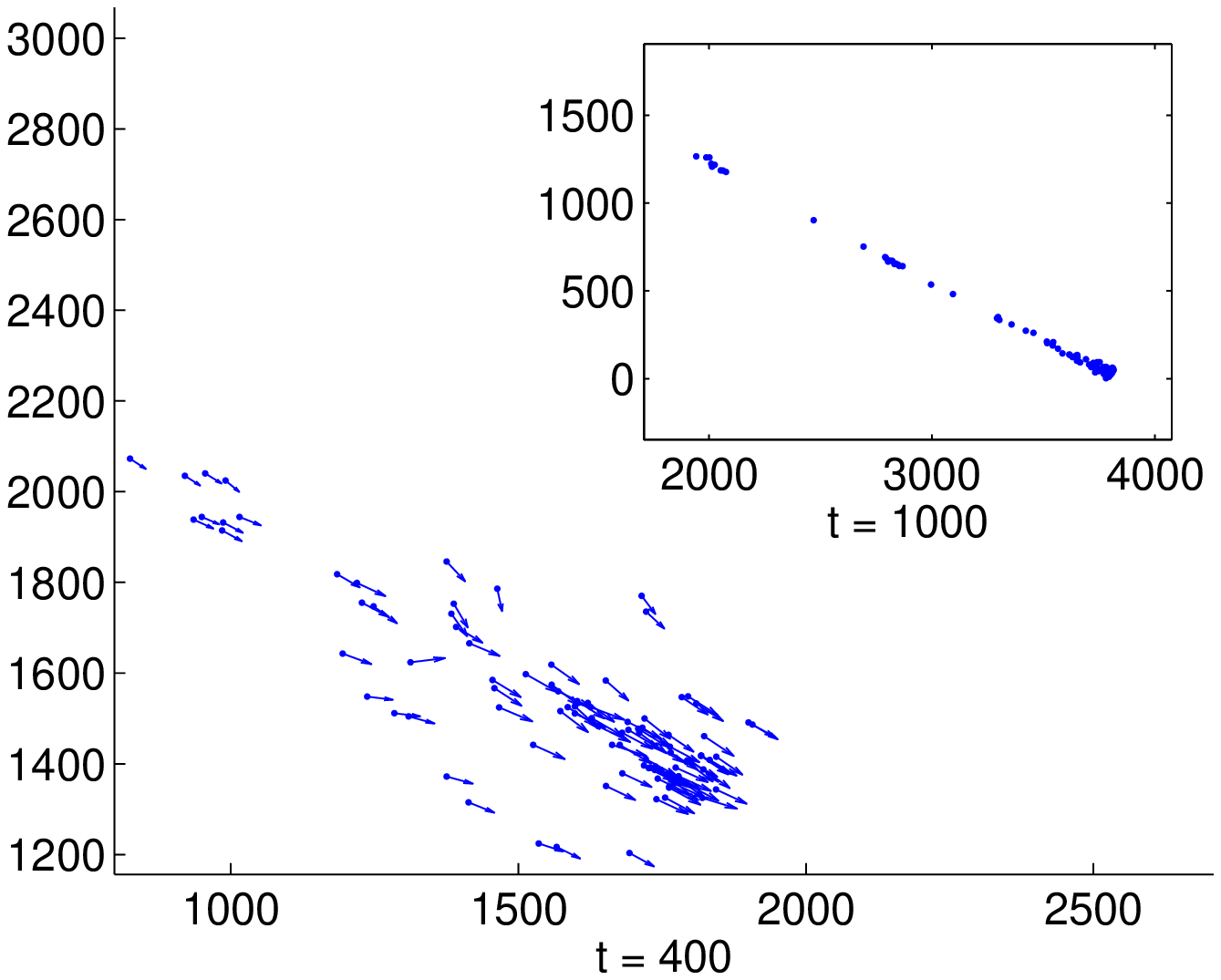} }
  \caption{Simulations with $M=50$, $r_c=500$, $\epsilon=0.5$. Random initial orientations in (a), snapshot at $t=400$. Initially aligned orientations in (b), snapshots at $t=400$ and $t=1000$  (enhanced online: sim4a.mp4 , sim4b.mp4) }
  \label{fig:figure4}
\end{figure}

After a long vortex simulation, the group might stabilize into a loosely cohesive unit that resembles a comet: a dense core followed by a tail of stranded individuals. This comet-like behavior can be obtained easier in a different simulation, by using similar parameters but considering initially aligned orientations instead. Figure 4(b) shows such an illustration of how the dense core is formed early on, and how the following individuals form a ``line" which lags behind, resembling a comet and its dissipating tail.   

Under the same conditions as the comet dynamics (high \emph{M} and aligned orientations), decreasing the initial density produces an interesting behavioral transition. Instead of going from uncertainty to order as in the vortex simulation (initially oscillating movement that later stabilizes), initially aligned individuals will later on reach opposing directions in the core as they try to attract each other, and form a counter-rotating center reminiscent of the vortex previously described: see Figure 5. Once the second state is reached, the population loses its initial global direction, and is anchored around the vortex center which can slightly move around randomly. The overall behavior can be described by saying that what would have been the core of the ``comet" becomes a vortex instead, causing oscillations. From our observations, the system consistently has this initial transition from order to uncertainty under these conditions, and it usually does not stabilize into global directed movement. We will denote this state as the semi-vortex. \\

\begin{figure}[htb]
  \centering  
  \includegraphics[scale=0.52]{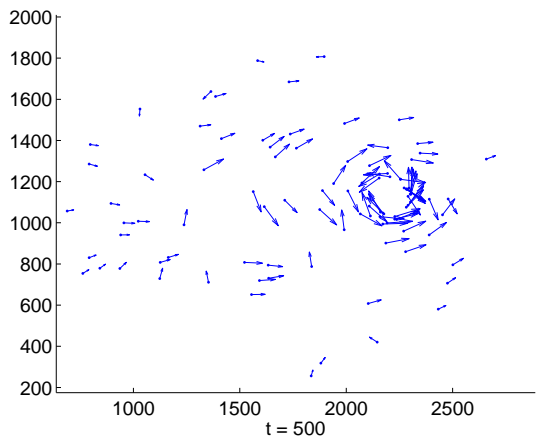} 
  \caption{From order to uncertainty (semi-vortex): a simulation with $M=50$, $r_c=1000$, $\epsilon=0.5$, and initially aligned orientations, at $t=500$ (enhanced online: sim5.mp4)}
 \label{fig:figure5}
\end{figure}

Changing to high density simulations, we observed interesting ``convergence" behaviors for some parameter ranges and alignments. Basically, two different groups are formed initially, resulting from the short-range repulsion at high densities, but later on they converge smoothly and re-organize into a single unit. This behavior was observed very frequently for $M=50$ at very high densities ($r_c$ between 10 and 50), and random initial orientations; though it was more consistent if aligning each half of the population with a $\pi/2$ angle difference. Figure 6(a) shows a snapshot of a simulation that exhibited this convergence, which results from a weak interaction still influencing each group due to the large value of \emph{M}, and causing a future re-grouping. 

\begin{figure}[htb]
  \centering
  \subfloat[Convergence]{\label{fig:figure6a}\includegraphics[scale=0.52]{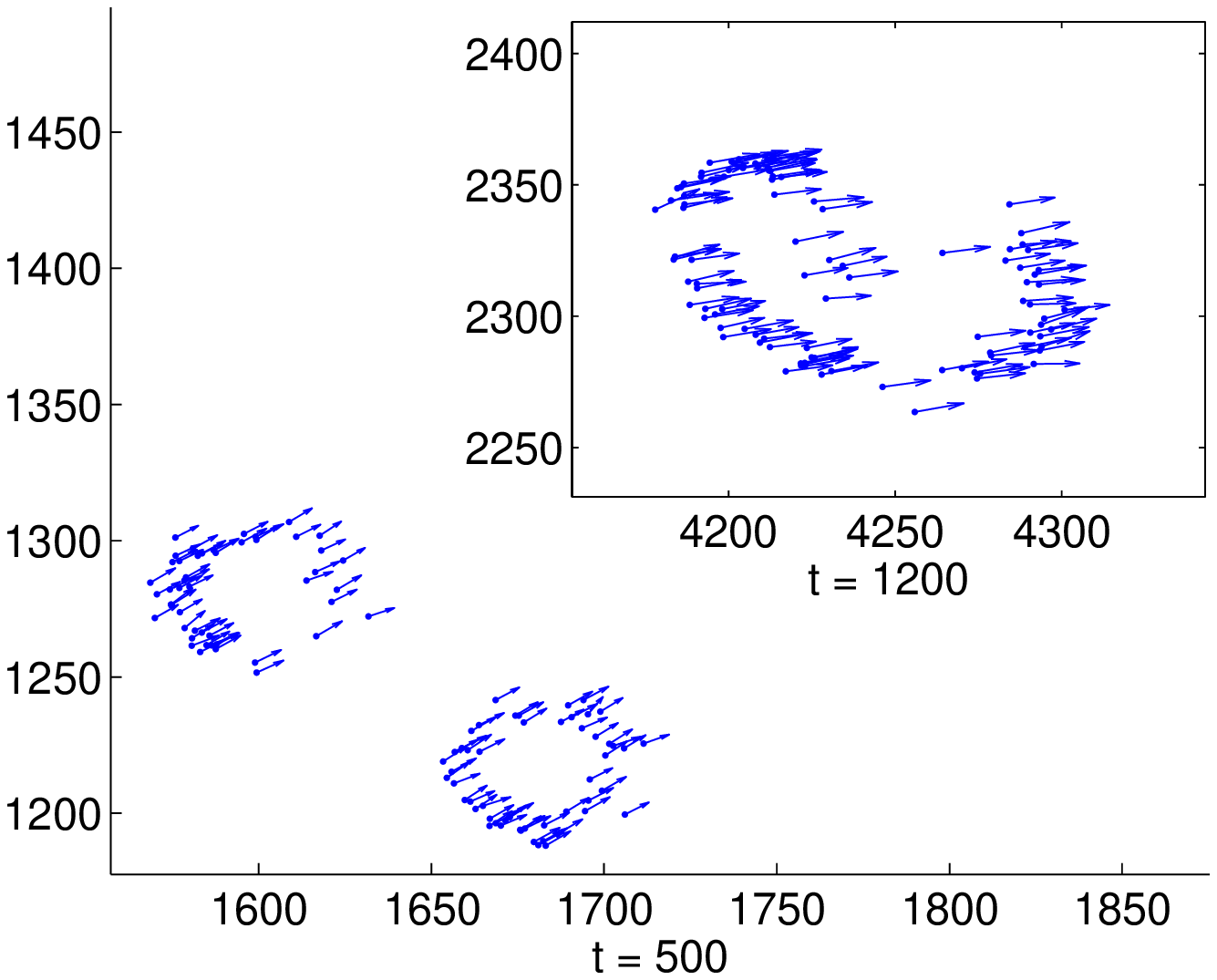}}             
  \hspace{0.1cm}   
  \subfloat[Tight cohesion]{\label{fig:figure6b}\includegraphics[scale=0.52]{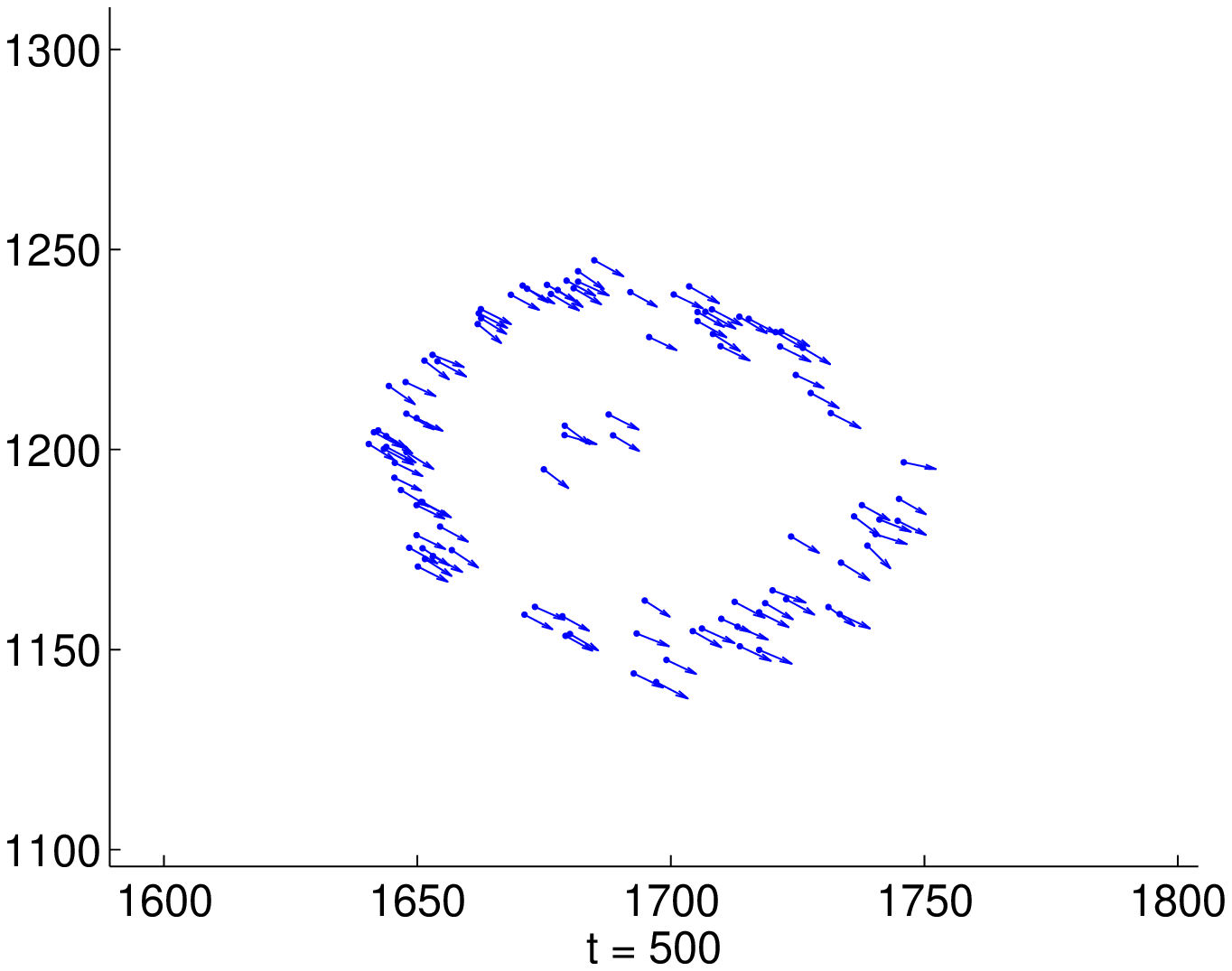} }
  \caption{Simulations with $M=50$, $r_c=50$, $\epsilon=0.5$. Divided initial orientations in (a), half the population pointing at 0 and the other half at $\pi/2$, snapshots at $t=500$ and $t=1200$. Initially aligned orientations in (b), snapshot at $t=500$ (enhanced online: sim6a.mp4 , sim6b.mp4)}
  \label{fig:figure6}
\end{figure}

Finally, for the similar parameters but with aligned orientations, the individuals form a compact cohesive unit and move together, as can be observed in Figure 6(b). Of interest here is that the individuals move within the pattern without affecting the shape or the global direction of the group, closely resembling the behavior of a bee swarm. This is likely due to the short-range repulsion and noise, creating sensitive deviations to a congestive formation. This behavior is consistent with the ``many wrongs" principle \cite{Simons2004}, in the sense that the many errors, deviations, or noise, within the individuals, are averaged out to form a single cohesive global direction which influences the navigation of the whole group. 

In order to give a better general picture of how changing density and speed affect the dynamics, we performed extensive manual visualization on simulations and classified them according to some of the behaviors just described. For this task, the initial radius $r_c$ was varied from 100 to 1000 in intervals of 20, and \emph{M} was spanned from 5 to 50 in unit intervals; thus giving 46x46 different parameter scenarios. These particular ranges were chosen because they were enough to span the significant behaviors in the model. To decisively associate a particular parameter setting to a behavior, we decided that at least four out of five of its simulations must show the behavior.  The cases with two possible behaviors with similar probabilities, were classified as ``variable behavior". The results from our visualizations can be seen in Figure 7, for both random and aligned initial orientations. We must emphasize that due to the manual visualization in the extensive number of simulations considered (large number of parameter settings, and the repetitions to ensure proper classification), the behavior zones are only approximate. Nevertheless, they give a good picture of how certain behaviors can be obtained and the phase transitions by changing either density or the number of interacting neighbors.           

\begin{figure}[htb]
  \centering
  \subfloat{\includegraphics[scale=0.60]{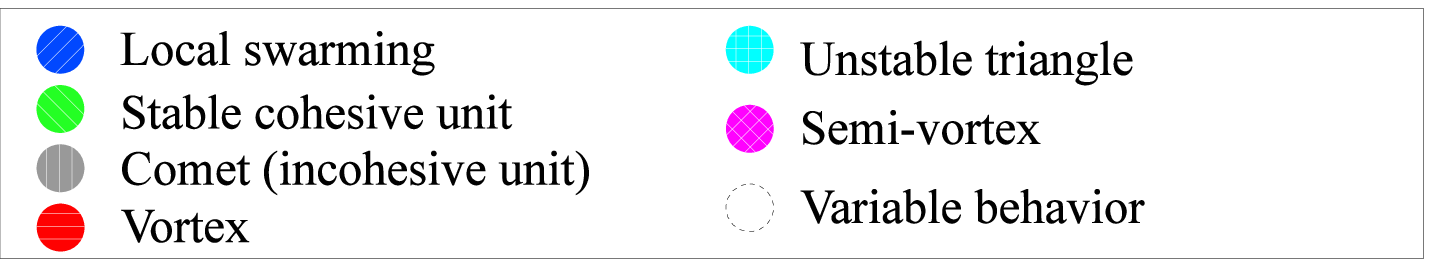}} \\
  \vspace{0.1cm}     
 \addtocounter{subfigure}{-1}
  \subfloat[Initially random orientations]{\label{fig:figure7a}\includegraphics[scale=0.52]{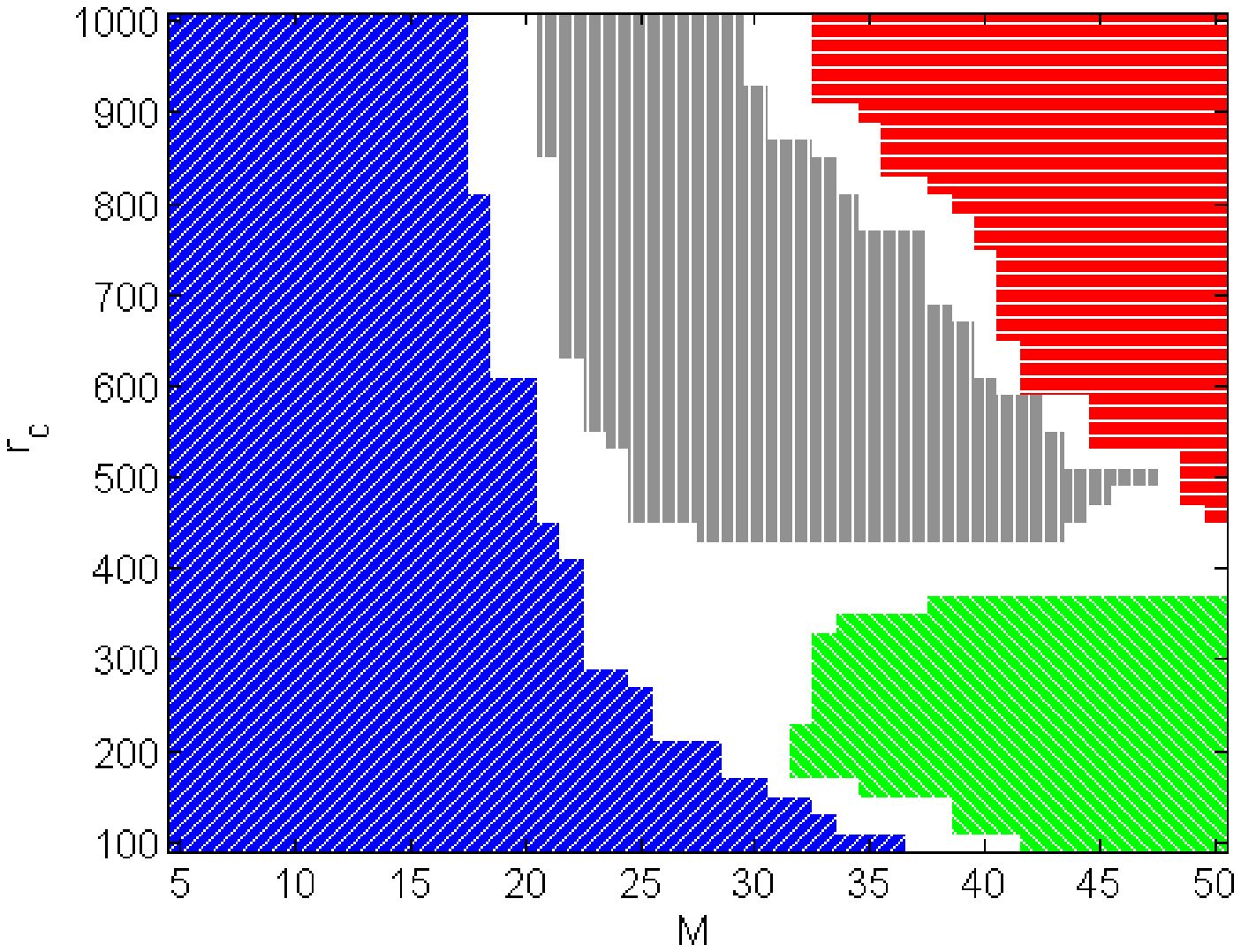}}             
  \hspace{0.1cm}   
  \subfloat[Initially aligned orientations]{\label{fig:figure7b}\includegraphics[scale=0.52]{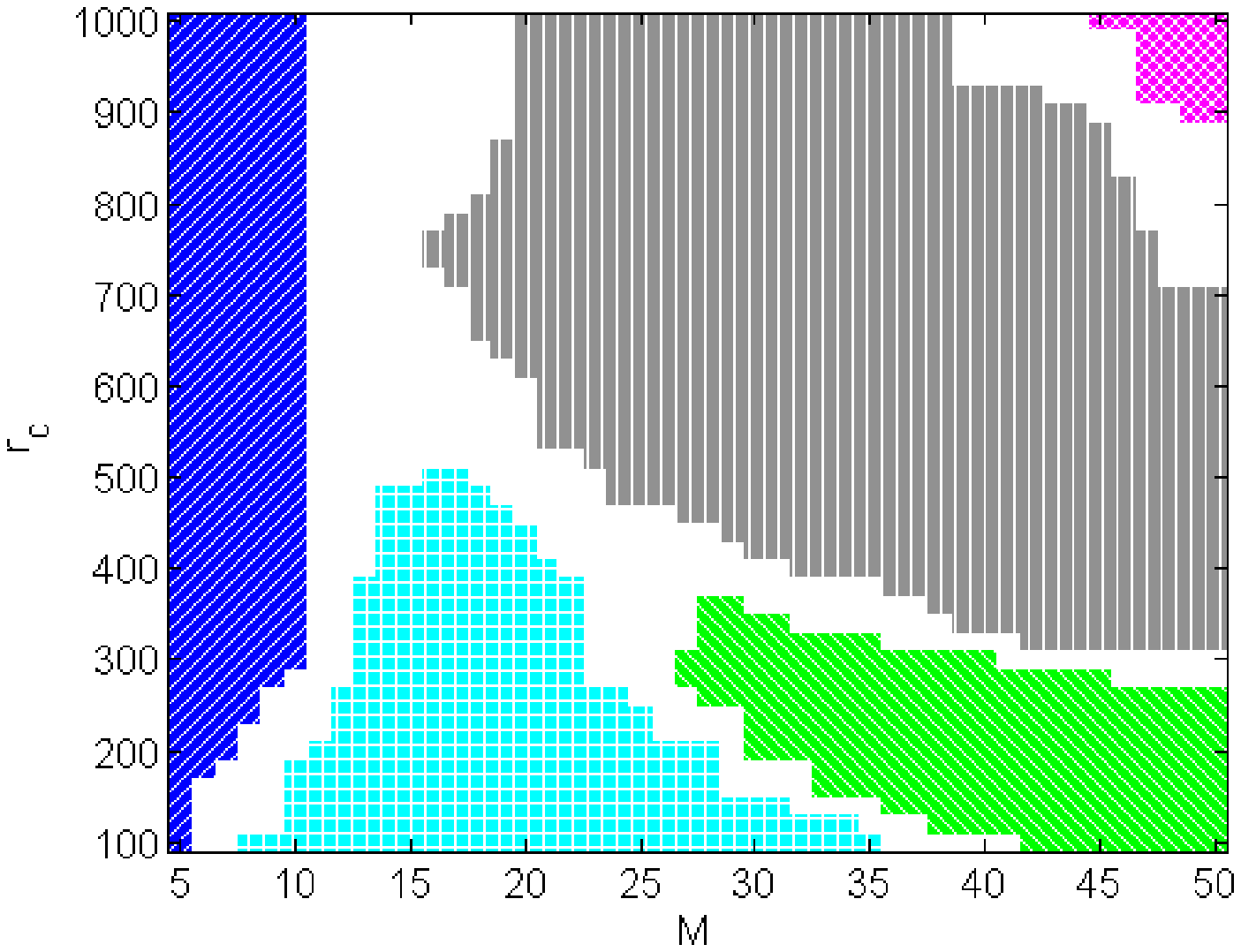} }
  \caption{Rough sketches of behavior zones for different initial densities (radius $r_c$) and interacting neighbors (\emph{M}); with a resolution of 46x46 parameter settings. All simulations had $N=100$ and $\epsilon=0.5$}
  \label{fig:figure7}
\end{figure}

Figure 7(a) shows that for initially random orientations, a transition from comet-like behavior to vortex can occur by increasing \emph{M} at low densities ($r_c > 450$). The same transition occurs by fixing \emph{M} near 40 and increasing $r_c$ at low densities. Cohesive behavior is associated with higher densities, and its change to either comet-like behavior or a vortex, depends of the value of \emph{M} when decreasing density; again a larger \emph{M} will cause vortex-like behavior. When considering initially aligned orientations in Figure 7(b), we have the change between unstable global movement to a cohesive unit by either increasing \emph{M} at high densities, or decreasing the density for values of \emph{M} near 30. The latter implies that higher densities do not necessarily imply best cohesion. The semi-vortex behavior only occupies the far corner region of low density and high \emph{M}, which tells that it will only emerge in cases of high separation and high degree of interaction. As a final but very interesting note, it is surprising to see that for $r_c$ near 350 and large \emph{M}, initially random orientations give cohesive behavior, but aligned orientations produce comet-like behavior with stranded individuals. This occurs in the latter because the initial equal orientation in the group causes the individuals at the front to synchronize fast as a single unit, while the rear part of the population struggles to keep up due to the weak attraction and low speeds for longer separations (see Figure 1 curves). This is actually the essence of the comet-like behavior, since the weak attraction is what causes the ``tail", as can be seen in figure 4(b). When using random initial orientations, the longer initial synchronization time actually helps the group stay together, and self-organize into a stable structure. It may be that the shapes formed by an initially disordered state are more stable than the ones formed with equal initial velocities, which underlines the complexity of the model dynamics and their structure.

\subsection{Quantitative measures of order}

Finding numerical measures that can quantify all the behaviors we have presented is a difficult task, and this is why we presented a phase diagram based on manual visualization. Nevertheless, we now use two order parameters to identify the global alignment and ``vorticity" in the system. First of all, the magnitude of the average velocity of the system at a given time interval:
\begin{equation}
v_m(t)=\left \| \frac{1}{N} \sum_{i=1}^{N} v_i(t) \right \|
\end{equation}               
is a good measure of the global order or direction of the system. A disordered system with individuals moving randomly should give a near zero $v_m(t)$ because the velocities cancel out, while an aligned swarm results in $v_m(t)$ equal to the average global speed of the group. The same or very similar measures have been used in previous studies for the same purpose \cite{Vicsek1995,Couzin2002}. To quantify the ``vorticity" of the swarm, i.e. how close it is to a vortex state, we measure the average absolute value of the angular velocities of the swarm at a given time:
\begin{equation}
\omega_m(t)= \frac{1}{N} \sum_{i=1}^{N} |\omega_i(t)|
\end{equation}  
which tells us how much the individuals are rotating around the center of mass of the swarm. We chose the average absolute value ($\frac{1}{N} \sum_{i=1}^{N} |\omega_i(t)|$) instead of the absolute value of the average ($\left | \frac{1}{N} \sum_{i=1}^{N}\omega_i(t) \right |$) of the angular velocities, in order to be able to identify the counter-rotating vortices, since the angular movement of many individuals would be opposite and cancel out with the latter measure, giving theoretically a low value which would make it difficult to distinguish from a cohesive state. In general, we expect $\omega_m(t)$ to be close to zero for an aligned noise-less flock heading in the same direction, but obviously greater for a system in a vortex state (regardless of the direction), with particles rotating around. The angular velocities $\omega_i(t)$ are estimated by taking phase differences: $\omega_i(t)=\theta_i(t)-\theta_i(t-1)$, where $\theta_i(t)$ is calculated from the x-axis to the relative position of individual \emph{i} to the mean position of the group at time \emph{t}: $\textbf{x}_i(t)-\langle \textbf{x}(t) \rangle$. 

\begin{figure}[t]
  \centering  
  \includegraphics[scale=0.50]{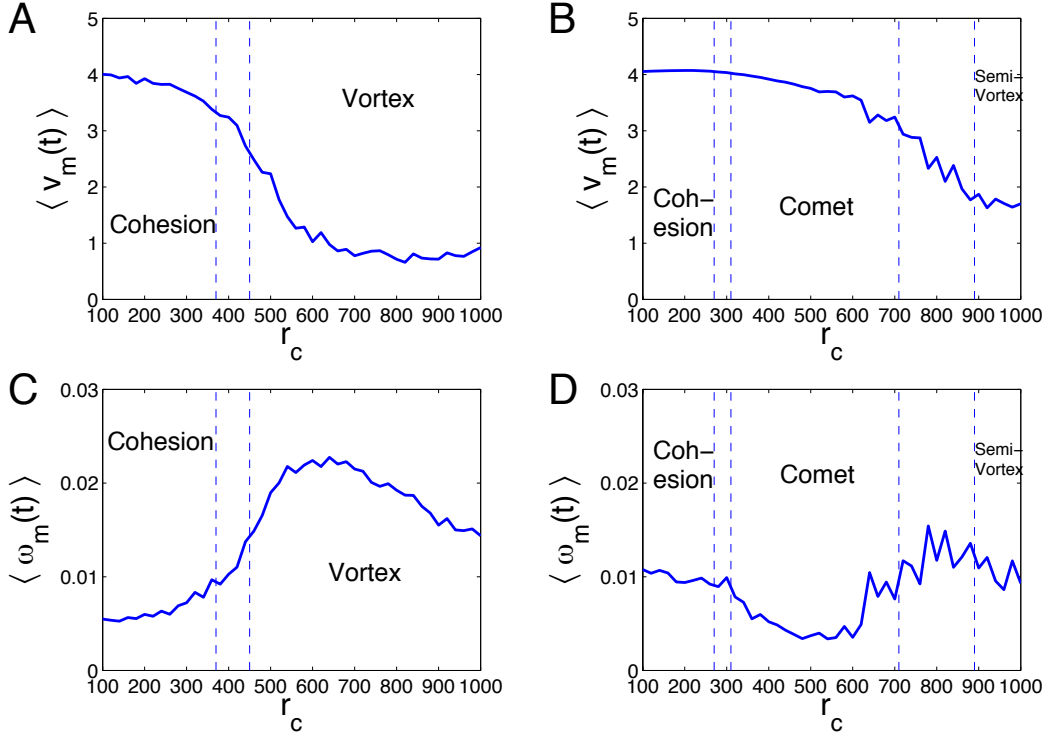} 
  \caption{Averaged $v_m(t)$ and $\omega_m(t)$ values for 50 simulations of 1000 time intervals at different initial densities (radius $r_c$) and $M=50$. Plots (a) and (c) correspond to random initial orientations, while (b) and (d) to aligned initial orientations. The transitions from a cohesive state to a vortex can be seen by the decrease of global direction $v_m(t)$, and increase of $\omega_m(t)$. The vertical lines correspond to the manual observations from Figure 7.  }
 \label{fig:figure8}
\end{figure}

With the purpose of verifying numerically the transition from a cohesive swarm to a vortex, we fix $M=50$ and span the same region of $r_c$ space as considered in the previous subsection, since we can see in Figure 7 that $M=50$ is a good enough value to cover the transition between these behaviors. For each $r_c$ value from 100 to 1000 (with increments of 20), we averaged $v_m(t)$ and $\omega_m(t)$ over 50 simulations and 1000 time intervals. Figure 8(a) shows that for random initial alignments, $v_m$ captures the loss of global direction as we go from the cohesive to the vortex state at around $r_c=400$. The increase in vorticity can be seen in Figure 8(c) for $\omega_m$, but interestingly, we see a drop near $r_c=1000$, and this implies that at lower densities the vortex loses force because some individuals do not interact as much. For initially aligned individuals, Figure 8(b) shows how the decrease in global direction is less strong, but decreases more significantly when we approach the transition from the comet to the semi-vortex state. This tells us that as we leave the comet state, the swarm loses global direction and the uncertainty in the system increases. Again we confirm this in Figure 8(d), with a very interesting reaction of $\omega_m$ as we approach the buffer area between the comet and the semi-vortex. The vorticity measure slightly increases but becomes very noisy as we approach $r_c=700$, and this implies that the global behavior is very uncertain in this parameter region.

\begin{figure}[htb]
  \centering
  \subfloat[Global direction]{\label{fig:figure9a}\includegraphics[scale=0.52]{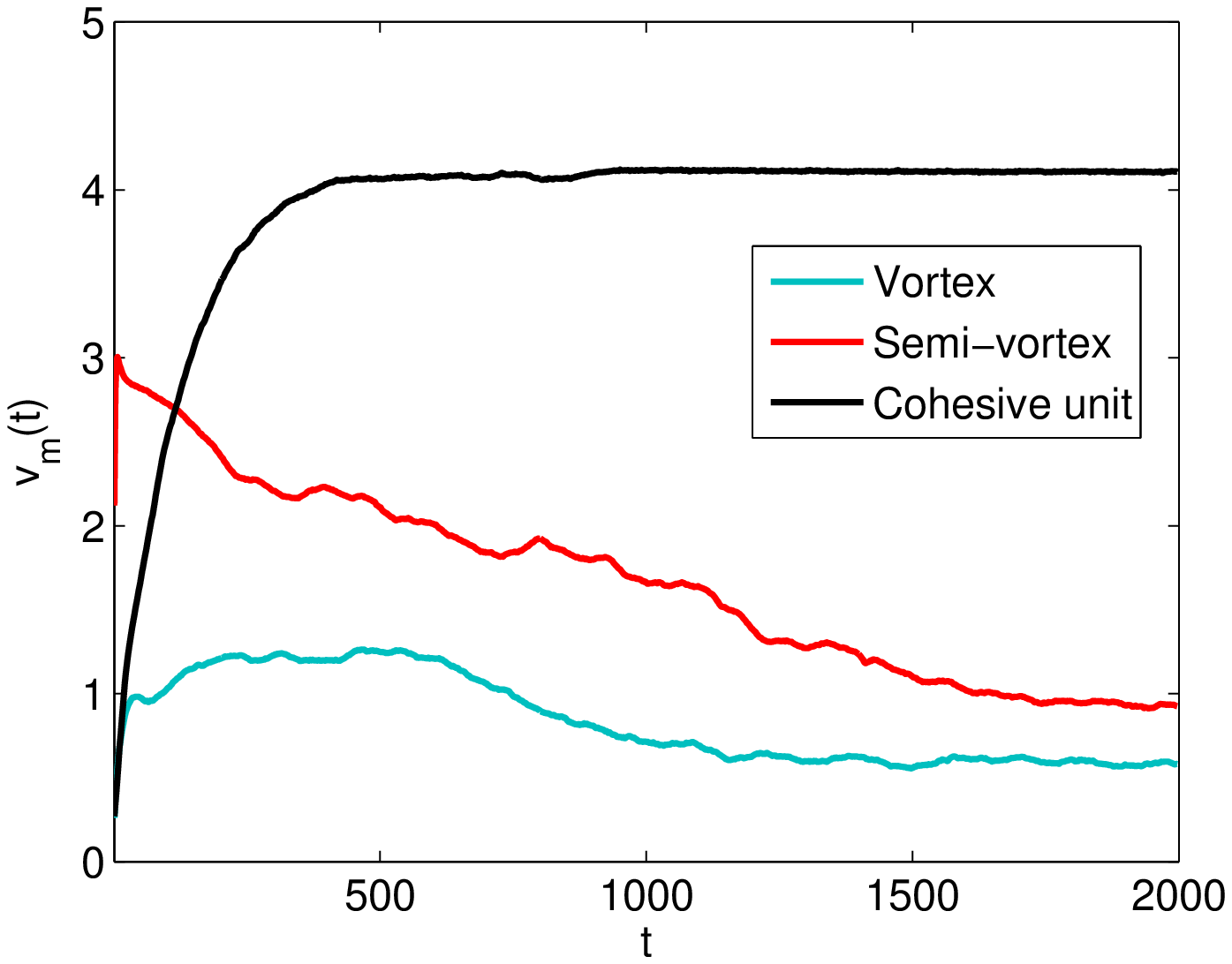}}             
  \subfloat[Vorticity]{\label{fig:figure9b}\includegraphics[scale=0.52]{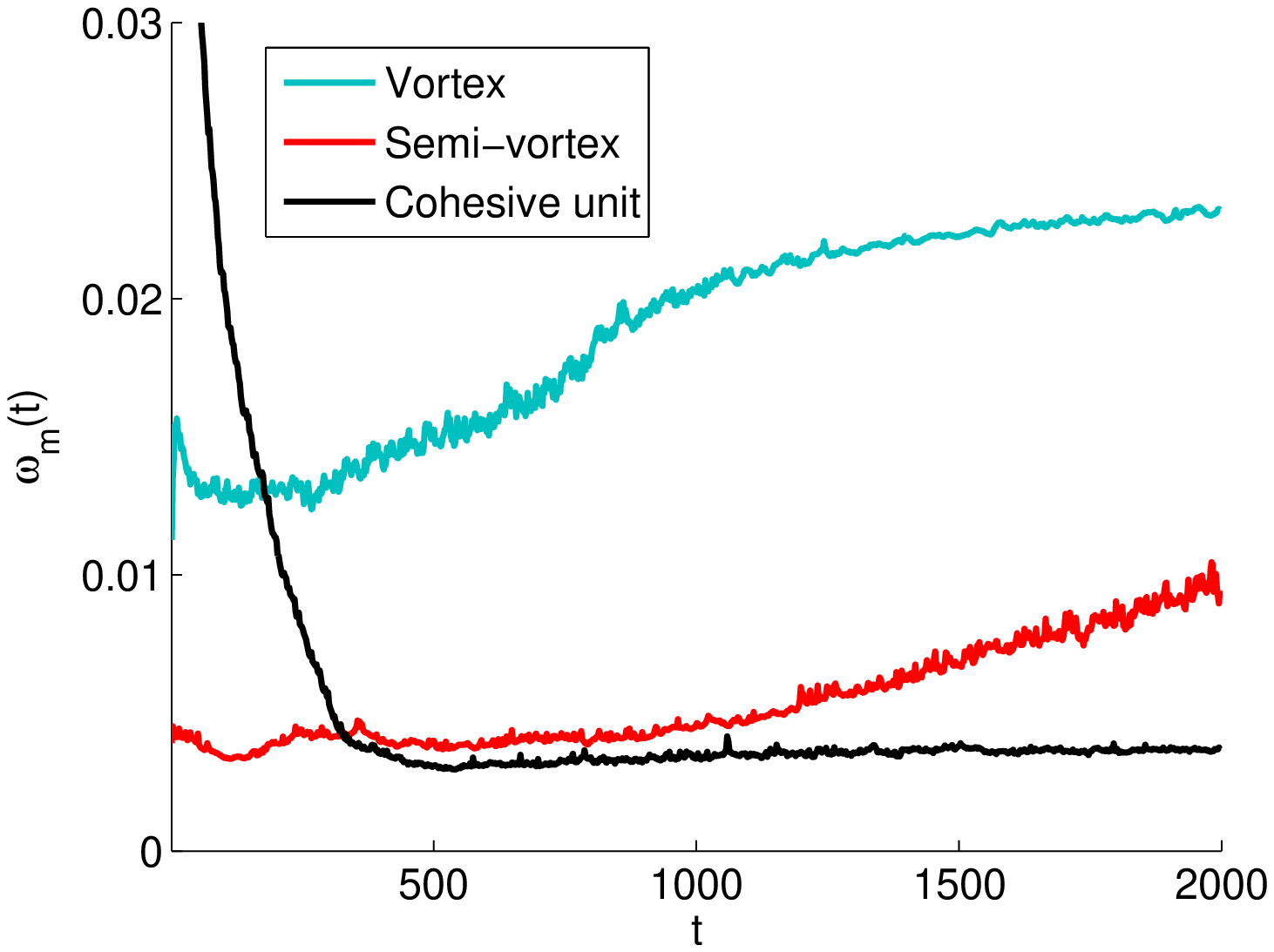} }
  \caption{Time courses of $v_m(t)$ and $\omega_m(t)$ for three scenarios ($M=50$): cohesive unit (random initial orientations, $r_c=200$), vortex (random initial orientations, $r_c=800$) and semi-vortex (aligned initial orientations, $r_c=950$). The time series were averaged over 50 simulations each.}
  \label{fig:figure9}
\end{figure}

In order to characterize the time evolution of the phase transitions, Figure 9 shows the averaged time courses of $v_m(t)$ and $\omega_m(t)$ for simulations that result in cohesive, vortex and semi-vortex states. In Figure 9(a) we can see how both the cohesive and vortex simulations start at disordered states (low $v_c(t)$), with the former steadily self-organizing into a cohesive flock and a stable $v_c(t)$, and the latter reaching no consensus. The semi-vortex runs start with an initially aligned group (high $v_c(t)$) which loses its direction as time advances. Figure 9(b) shows how the cohesive simulations start with a very high vorticity rate in the synchronization stage, and later it is decreased as the swarm aligns. Both the vortex and semi-vortex simulations have increases of $\omega_m(t)$ as time advances, with higher values for the former. Interestingly, the measures tell us how the cohesive simulations start in a disordered state and later self-organize into a stable group, as well as how for semi-vortex simulations, an initially aligned swarm evolves into an uncertain state that loses its global direction and results in some individuals forming a smaller vortex. In general, the transitions in Figure 9 are consistent with our observations of the previous subsection. These final illustrations show us how both order parameters provide a clear description of the global alignment and vorticity of the population, and can characterize the transitions that occur.

\section{Conclusions}

We must emphasize that roughly the same behaviors as the ones described were also observed qualitatively for swarms of smaller or larger sizes. As should be expected for different population sizes, the parameters required to get a desired behavior are different, due to the difference in density properties and neighbor interactions. Nevertheless, even though we tried to summarize the most significant behaviors that were found in our extensive simulations, there might be other relevant behaviors that we have not discovered and require more exhaustive computation or a deeper mathematical analysis. Still, we present our results as an initial taste of the wide variety of realistic behaviors that can be obtained from our simple model based on attraction and velocity distributions abstracted from pigeon flock data.

In our previous work \cite{Kattas2011b}, we built a computational model from experimental data and demonstrated that it is capable of performing accurate simulations of collective pigeon flights. In this paper we used a simple attraction/repulsion rule extracted from this model. We find that this rule is similar though significantly different from those rules used in the standard models of collective behavior \cite{Vicsek1995, Reynolds1987}. Our main motivation in this current work is to make a simpler abstracted naive model: a model based on this inferred rule, which is nonetheless tractable and capable of simulating a wide range of phenomena. We then simulated this naive model on a range of parameter values far broader than the original experimental system, and demonstrated it is capable of emulating a considerable variety of realistic collective behaviors that resemble natural phenomena.

Several models have individually reported similar behaviors as the ones we have described. To name a few, Couzin et al. documented swarming, cohesive, and milling (vortex) dynamics by varying the interaction zones of individuals \cite{Couzin2002}. Chuang et al. \cite{Chuang2007} and Touma et al. \cite{Touma2010} have shown counter-rotating vortices by changing the attraction and repulsion strengths of their models. Similar dynamics to our semi-vortex state, where initially aligned individuals later produce oscillations, have been described by Erdmann et al. \cite{Erdmann2005}, using a model which only considers attraction between individuals and noise. The strength of our model lies in the fact that by following a rule of the basic attraction and repulsion mechanisms extracted from experimental data, we can generate this wide range of interesting complex behaviors. In addition, to our knowledge, the comet dynamics (Figure 4(b)) have not been documented before. This behavior occurs because the weak attractions at longer separations in our curve (see Figure 1(a)) cause the more distant individuals to lag behind the core in low density scenarios, and thus are a result of using our experimentally motivated attraction/repulsion dynamics. Pattern transitions are also now being considered in models for collective dynamics \cite{Cheng2011}, and furthermore, our simulations also show pattern formations such as loose triangles (Figure 2(b)) and other shapes (Figure 3(b)), making the range of behaviors even larger. Finally, we stress that our model is capable of producing complex self-organized behavior without a global navigational mechanism such as a ``leader", external perturbations from the terrain, or a particularly fixed particle distribution; and this is achieved by only changing the initial density conditions and the number of interacting neighbors. 

Also worth mentioning is the fact that the stable behaviors observed in our model show higher densities at the shape borders than in the center. This type of density distribution is consistent with what was measured in a recent investigation of starling flocks from real data \cite{Ballerini2008a}. The existence of optimal density conditions, implying that initially high densities do not guarantee best cohesion, which was observed in our previous study \cite{Kattas2011b}, was reconfirmed with our model. Our simulations are also consistent with the ``many wrongs" principle \cite{Simons2004}, since in many of the runs that showed cohesive behavior, the individual errors of the individuals did not affect the global navigational direction of the whole swarm, and it showed strong, cohesive, and stable motion. We believe that what is presented in this paper is simply the tip of the iceberg of what could be obtained by building mathematical models which attempt to follow or preserve rules extracted from real experimental data, especially in collective systems.


%
%

%

\begin{acknowledgments}
GDK is currently supported by the Hong Kong PHD Fellowship Scheme (HKPFS) from the Research Grants Council (RGC) of Hong Kong. XKX is currently supported by the PolyU Postdoctoral Fellowships Scheme (G-YX4A) and the Research Grants Council of Hong Kong (BQ19H). XKX also acknowledges the Natural Science Foundation of China (61004104, 61104143).
\end{acknowledgments}


\begin{thebibliography}{27}%
\makeatletter
\providecommand \@ifxundefined [1]{%
 \@ifx{#1\undefined}
}%
\providecommand \@ifnum [1]{%
 \ifnum #1\expandafter \@firstoftwo
 \else \expandafter \@secondoftwo
 \fi
}%
\providecommand \@ifx [1]{%
 \ifx #1\expandafter \@firstoftwo
 \else \expandafter \@secondoftwo
 \fi
}%
\providecommand \natexlab [1]{#1}%
\providecommand \enquote  [1]{``#1''}%
\providecommand \bibnamefont  [1]{#1}%
\providecommand \bibfnamefont [1]{#1}%
\providecommand \citenamefont [1]{#1}%
\providecommand \href@noop [0]{\@secondoftwo}%
\providecommand \href [0]{\begingroup \@sanitize@url \@href}%
\providecommand \@href[1]{\@@startlink{#1}\@@href}%
\providecommand \@@href[1]{\endgroup#1\@@endlink}%
\providecommand \@sanitize@url [0]{\catcode `\\12\catcode `\$12\catcode
  `\&12\catcode `\#12\catcode `\^12\catcode `\_12\catcode `\%12\relax}%
\providecommand \@@startlink[1]{}%
\providecommand \@@endlink[0]{}%
\providecommand \url  [0]{\begingroup\@sanitize@url \@url }%
\providecommand \@url [1]{\endgroup\@href {#1}{\urlprefix }}%
\providecommand \urlprefix  [0]{URL }%
\providecommand \Eprint [0]{\href }%
\providecommand \doibase [0]{http://dx.doi.org/}%
\providecommand \selectlanguage [0]{\@gobble}%
\providecommand \bibinfo  [0]{\@secondoftwo}%
\providecommand \bibfield  [0]{\@secondoftwo}%
\providecommand \translation [1]{[#1]}%
\providecommand \BibitemOpen [0]{}%
\providecommand \bibitemStop [0]{}%
\providecommand \bibitemNoStop [0]{.\EOS\space}%
\providecommand \EOS [0]{\spacefactor3000\relax}%
\providecommand \BibitemShut  [1]{\csname bibitem#1\endcsname}%
\let\auto@bib@innerbib\@empty
\bibitem [{\citenamefont {Okubo}\ and\ \citenamefont
  {Levin}(2001)}]{Okubo2001}%
  \BibitemOpen
  \bibfield  {author} {\bibinfo {author} {\bibfnamefont {A.}~\bibnamefont
  {Okubo}}\ and\ \bibinfo {author} {\bibfnamefont {S.~A.}\ \bibnamefont
  {Levin}},\ }\href@noop {} {\emph {\bibinfo {title} {Diffusion and Ecological
  Problems: Modern Perspectives}}}\ (\bibinfo  {publisher} {Springer Verlag},\
  \bibinfo {year} {2001})\BibitemShut {NoStop}%
\bibitem [{\citenamefont {Potts}(1984)}]{Potts1984}%
  \BibitemOpen
  \bibfield  {author} {\bibinfo {author} {\bibfnamefont {W.~K.}\ \bibnamefont
  {Potts}},\ }\href@noop {} {\bibfield  {journal} {\bibinfo  {journal}
  {Nature}\ }\textbf {\bibinfo {volume} {309}},\ \bibinfo {pages} {344}
  (\bibinfo {year} {1984})}\BibitemShut {NoStop}%
\bibitem [{\citenamefont {Couzin}\ \emph {et~al.}(2005)\citenamefont {Couzin},
  \citenamefont {Krause}, \citenamefont {Franks},\ and\ \citenamefont
  {Levin}}]{Couzin2005}%
  \BibitemOpen
  \bibfield  {author} {\bibinfo {author} {\bibfnamefont {I.~D.}\ \bibnamefont
  {Couzin}}, \bibinfo {author} {\bibfnamefont {J.}~\bibnamefont {Krause}},
  \bibinfo {author} {\bibfnamefont {N.~R.}\ \bibnamefont {Franks}}, \ and\
  \bibinfo {author} {\bibfnamefont {S.~A.}\ \bibnamefont {Levin}},\ }\href@noop
  {} {\bibfield  {journal} {\bibinfo  {journal} {Nature}\ }\textbf {\bibinfo
  {volume} {433}},\ \bibinfo {pages} {513} (\bibinfo {year}
  {2005})}\BibitemShut {NoStop}%
\bibitem [{\citenamefont {Conradt}\ and\ \citenamefont
  {Roper}(2003)}]{Conradt2003}%
  \BibitemOpen
  \bibfield  {author} {\bibinfo {author} {\bibfnamefont {L.}~\bibnamefont
  {Conradt}}\ and\ \bibinfo {author} {\bibfnamefont {T.~J.}\ \bibnamefont
  {Roper}},\ }\href@noop {} {\bibfield  {journal} {\bibinfo  {journal}
  {Nature}\ }\textbf {\bibinfo {volume} {421}},\ \bibinfo {pages} {155}
  (\bibinfo {year} {2003})}\BibitemShut {NoStop}%
\bibitem [{\citenamefont {Simons}(2004)}]{Simons2004}%
  \BibitemOpen
  \bibfield  {author} {\bibinfo {author} {\bibfnamefont {A.~M.}\ \bibnamefont
  {Simons}},\ }\href@noop {} {\bibfield  {journal} {\bibinfo  {journal} {Trends
  in Ecology \& Evolution}\ }\textbf {\bibinfo {volume} {19}},\ \bibinfo
  {pages} {453} (\bibinfo {year} {2004})}\BibitemShut {NoStop}%
\bibitem [{\citenamefont {Codling}, \citenamefont {Pitchford},\ and\
  \citenamefont {Simpson}(2007)}]{Codling2007}%
  \BibitemOpen
  \bibfield  {author} {\bibinfo {author} {\bibfnamefont {E.~A.}\ \bibnamefont
  {Codling}}, \bibinfo {author} {\bibfnamefont {J.~W.}\ \bibnamefont
  {Pitchford}}, \ and\ \bibinfo {author} {\bibfnamefont {S.~D.}\ \bibnamefont
  {Simpson}},\ }\href@noop {} {\bibfield  {journal} {\bibinfo  {journal}
  {Ecology}\ }\textbf {\bibinfo {volume} {88}},\ \bibinfo {pages} {1864}
  (\bibinfo {year} {2007})}\BibitemShut {NoStop}%
\bibitem [{\citenamefont {Nagy}\ \emph {et~al.}(2010)\citenamefont {Nagy},
  \citenamefont {Akos}, \citenamefont {Biro},\ and\ \citenamefont
  {Vicsek}}]{Nagy2010}%
  \BibitemOpen
  \bibfield  {author} {\bibinfo {author} {\bibfnamefont {M.}~\bibnamefont
  {Nagy}}, \bibinfo {author} {\bibfnamefont {Z.}~\bibnamefont {Akos}}, \bibinfo
  {author} {\bibfnamefont {D.}~\bibnamefont {Biro}}, \ and\ \bibinfo {author}
  {\bibfnamefont {T.}~\bibnamefont {Vicsek}},\ }\href@noop {} {\bibfield
  {journal} {\bibinfo  {journal} {Nature}\ }\textbf {\bibinfo {volume} {464}},\
  \bibinfo {pages} {890} (\bibinfo {year} {2010})}\BibitemShut {NoStop}%
\bibitem [{\citenamefont {Ballerini}\ \emph
  {et~al.}(2008{\natexlab{a}})\citenamefont {Ballerini}, \citenamefont
  {Cabibbo}, \citenamefont {Candelier}, \citenamefont {Cavagna}, \citenamefont
  {Cisbani}, \citenamefont {Giardina}, \citenamefont {Lecomte}, \citenamefont
  {Orlandi}, \citenamefont {Parisi}, \citenamefont {Procaccini}, \citenamefont
  {Viale},\ and\ \citenamefont {Zdravkovic}}]{Ballerini2008}%
  \BibitemOpen
  \bibfield  {author} {\bibinfo {author} {\bibfnamefont {M.}~\bibnamefont
  {Ballerini}}, \bibinfo {author} {\bibfnamefont {N.}~\bibnamefont {Cabibbo}},
  \bibinfo {author} {\bibfnamefont {R.}~\bibnamefont {Candelier}}, \bibinfo
  {author} {\bibfnamefont {A.}~\bibnamefont {Cavagna}}, \bibinfo {author}
  {\bibfnamefont {E.}~\bibnamefont {Cisbani}}, \bibinfo {author} {\bibfnamefont
  {I.}~\bibnamefont {Giardina}}, \bibinfo {author} {\bibfnamefont
  {V.}~\bibnamefont {Lecomte}}, \bibinfo {author} {\bibfnamefont
  {A.}~\bibnamefont {Orlandi}}, \bibinfo {author} {\bibfnamefont
  {G.}~\bibnamefont {Parisi}}, \bibinfo {author} {\bibfnamefont
  {A.}~\bibnamefont {Procaccini}}, \bibinfo {author} {\bibfnamefont
  {M.}~\bibnamefont {Viale}}, \ and\ \bibinfo {author} {\bibfnamefont
  {V.}~\bibnamefont {Zdravkovic}},\ }\href@noop {} {\bibfield  {journal}
  {\bibinfo  {journal} {Proceedings of the National Academy of Sciences}\
  }\textbf {\bibinfo {volume} {105}},\ \bibinfo {pages} {1232} (\bibinfo {year}
  {2008}{\natexlab{a}})}\BibitemShut {NoStop}%
\bibitem [{\citenamefont {Reynolds}(1987)}]{Reynolds1987}%
  \BibitemOpen
  \bibfield  {author} {\bibinfo {author} {\bibfnamefont {C.~W.}\ \bibnamefont
  {Reynolds}},\ }\href@noop {} {\bibfield  {journal} {\bibinfo  {journal}
  {SIGGRAPH Comput. Graph.}\ }\textbf {\bibinfo {volume} {21}},\ \bibinfo
  {pages} {25} (\bibinfo {year} {1987})}\BibitemShut {NoStop}%
\bibitem [{\citenamefont {Vicsek}\ \emph {et~al.}(1995)\citenamefont {Vicsek},
  \citenamefont {Czirok}, \citenamefont {Ben-Jacob}, \citenamefont {Cohen},\
  and\ \citenamefont {Shochet}}]{Vicsek1995}%
  \BibitemOpen
  \bibfield  {author} {\bibinfo {author} {\bibfnamefont {T.}~\bibnamefont
  {Vicsek}}, \bibinfo {author} {\bibfnamefont {A.}~\bibnamefont {Czirok}},
  \bibinfo {author} {\bibfnamefont {E.}~\bibnamefont {Ben-Jacob}}, \bibinfo
  {author} {\bibfnamefont {I.}~\bibnamefont {Cohen}}, \ and\ \bibinfo {author}
  {\bibfnamefont {O.}~\bibnamefont {Shochet}},\ }\href@noop {} {\bibfield
  {journal} {\bibinfo  {journal} {Phys. Rev. Lett.}\ }\textbf {\bibinfo
  {volume} {75}},\ \bibinfo {pages} {1226} (\bibinfo {year}
  {1995})}\BibitemShut {NoStop}%
\bibitem [{\citenamefont {Couzin}\ \emph {et~al.}(2002)\citenamefont {Couzin},
  \citenamefont {Krause}, \citenamefont {James}, \citenamefont {Ruxton},\ and\
  \citenamefont {Franks}}]{Couzin2002}%
  \BibitemOpen
  \bibfield  {author} {\bibinfo {author} {\bibfnamefont {I.~D.}\ \bibnamefont
  {Couzin}}, \bibinfo {author} {\bibfnamefont {J.}~\bibnamefont {Krause}},
  \bibinfo {author} {\bibfnamefont {R.}~\bibnamefont {James}}, \bibinfo
  {author} {\bibfnamefont {G.~D.}\ \bibnamefont {Ruxton}}, \ and\ \bibinfo
  {author} {\bibfnamefont {N.~R.}\ \bibnamefont {Franks}},\ }\href
  {http://www.sciencedirect.com/science/article/pii/S0022519302930651}
  {\bibfield  {journal} {\bibinfo  {journal} {Journal of Theoretical Biology}\
  }\textbf {\bibinfo {volume} {218}},\ \bibinfo {pages} {1} (\bibinfo {year}
  {2002})}\BibitemShut {NoStop}%
\bibitem [{\citenamefont {Erdmann}, \citenamefont {Ebeling},\ and\
  \citenamefont {Mikhailov}(2005)}]{Erdmann2005}%
  \BibitemOpen
  \bibfield  {author} {\bibinfo {author} {\bibfnamefont {U.}~\bibnamefont
  {Erdmann}}, \bibinfo {author} {\bibfnamefont {W.}~\bibnamefont {Ebeling}}, \
  and\ \bibinfo {author} {\bibfnamefont {A.~S.}\ \bibnamefont {Mikhailov}},\
  }\href {http://link.aps.org/doi/10.1103/PhysRevE.71.051904} {\bibfield
  {journal} {\bibinfo  {journal} {Phys. Rev. E}\ }\textbf {\bibinfo {volume}
  {71}},\ \bibinfo {pages} {051904} (\bibinfo {year} {2005})}\BibitemShut
  {NoStop}%
\bibitem [{\citenamefont {Chuang}\ \emph {et~al.}(2007)\citenamefont {Chuang},
  \citenamefont {D'Orsogna}, \citenamefont {Marthaler}, \citenamefont
  {Bertozzi},\ and\ \citenamefont {Chayes}}]{Chuang2007}%
  \BibitemOpen
  \bibfield  {author} {\bibinfo {author} {\bibfnamefont {Y.-L.}\ \bibnamefont
  {Chuang}}, \bibinfo {author} {\bibfnamefont {M.~R.}\ \bibnamefont
  {D'Orsogna}}, \bibinfo {author} {\bibfnamefont {D.}~\bibnamefont
  {Marthaler}}, \bibinfo {author} {\bibfnamefont {A.~L.}\ \bibnamefont
  {Bertozzi}}, \ and\ \bibinfo {author} {\bibfnamefont {L.~S.}\ \bibnamefont
  {Chayes}},\ }\href
  {http://www.sciencedirect.com/science/article/pii/S016727890700156X}
  {\bibfield  {journal} {\bibinfo  {journal} {Physica D: Nonlinear Phenomena}\
  }\textbf {\bibinfo {volume} {232}},\ \bibinfo {pages} {33} (\bibinfo {year}
  {2007})}\BibitemShut {NoStop}%
\bibitem [{\citenamefont {Touma}, \citenamefont {Shreim},\ and\ \citenamefont
  {Klushin}(2010)}]{Touma2010}%
  \BibitemOpen
  \bibfield  {author} {\bibinfo {author} {\bibfnamefont {J.~R.}\ \bibnamefont
  {Touma}}, \bibinfo {author} {\bibfnamefont {A.}~\bibnamefont {Shreim}}, \
  and\ \bibinfo {author} {\bibfnamefont {L.~I.}\ \bibnamefont {Klushin}},\
  }\href {http://link.aps.org/doi/10.1103/PhysRevE.81.066106} {\bibfield
  {journal} {\bibinfo  {journal} {Phys. Rev. E}\ }\textbf {\bibinfo {volume}
  {81}},\ \bibinfo {pages} {066106} (\bibinfo {year} {2010})}\BibitemShut
  {NoStop}%
\bibitem [{\citenamefont {Raghib}, \citenamefont {Levin},\ and\ \citenamefont
  {Kevrekidis}(2010)}]{Raghib2010}%
  \BibitemOpen
  \bibfield  {author} {\bibinfo {author} {\bibfnamefont {M.}~\bibnamefont
  {Raghib}}, \bibinfo {author} {\bibfnamefont {S.}~\bibnamefont {Levin}}, \
  and\ \bibinfo {author} {\bibfnamefont {I.}~\bibnamefont {Kevrekidis}},\
  }\href {http://www.sciencedirect.com/science/article/pii/S0022519310001050}
  {\bibfield  {journal} {\bibinfo  {journal} {Journal of Theoretical Biology}\
  }\textbf {\bibinfo {volume} {264}},\ \bibinfo {pages} {893} (\bibinfo {year}
  {2010})}\BibitemShut {NoStop}%
\bibitem [{\citenamefont {Zhang}\ \emph {et~al.}(2010)\citenamefont {Zhang},
  \citenamefont {Wang}, \citenamefont {Chen}, \citenamefont {Su}, \citenamefont
  {Zhou},\ and\ \citenamefont {Zhou}}]{Zhang2010}%
  \BibitemOpen
  \bibfield  {author} {\bibinfo {author} {\bibfnamefont {H.-T.}\ \bibnamefont
  {Zhang}}, \bibinfo {author} {\bibfnamefont {N.}~\bibnamefont {Wang}},
  \bibinfo {author} {\bibfnamefont {M.~Z.}\ \bibnamefont {Chen}}, \bibinfo
  {author} {\bibfnamefont {R.-Q.}\ \bibnamefont {Su}}, \bibinfo {author}
  {\bibfnamefont {T.}~\bibnamefont {Zhou}}, \ and\ \bibinfo {author}
  {\bibfnamefont {C.}~\bibnamefont {Zhou}},\ }\href@noop {} {\bibfield
  {journal} {\bibinfo  {journal} {New Journal of Physics}\ }\textbf {\bibinfo
  {volume} {12}},\ \bibinfo {pages} {123025} (\bibinfo {year}
  {2010})}\BibitemShut {NoStop}%
\bibitem [{\citenamefont {Lukeman}, \citenamefont {Li},\ and\ \citenamefont
  {Edelstein-Keshet}(2010)}]{Lukeman2010}%
  \BibitemOpen
  \bibfield  {author} {\bibinfo {author} {\bibfnamefont {R.}~\bibnamefont
  {Lukeman}}, \bibinfo {author} {\bibfnamefont {Y.-X.}\ \bibnamefont {Li}}, \
  and\ \bibinfo {author} {\bibfnamefont {L.}~\bibnamefont {Edelstein-Keshet}},\
  }\href@noop {} {\bibfield  {journal} {\bibinfo  {journal} {Proceedings of the
  National Academy of Sciences}\ }\textbf {\bibinfo {volume} {107}},\ \bibinfo
  {pages} {12576} (\bibinfo {year} {2010})}\BibitemShut {NoStop}%
\bibitem [{\citenamefont {Eriksson}\ \emph {et~al.}(2010)\citenamefont
  {Eriksson}, \citenamefont {Nilsson~Jacobi}, \citenamefont {Nystrom},\ and\
  \citenamefont {Tunstrom}}]{Eriksson2010}%
  \BibitemOpen
  \bibfield  {author} {\bibinfo {author} {\bibfnamefont {A.}~\bibnamefont
  {Eriksson}}, \bibinfo {author} {\bibfnamefont {M.}~\bibnamefont
  {Nilsson~Jacobi}}, \bibinfo {author} {\bibfnamefont {J.}~\bibnamefont
  {Nystrom}}, \ and\ \bibinfo {author} {\bibfnamefont {K.}~\bibnamefont
  {Tunstrom}},\ }\href@noop {} {\bibfield  {journal} {\bibinfo  {journal}
  {Behavioral Ecology}\ }\textbf {\bibinfo {volume} {21}},\ \bibinfo {pages}
  {1106} (\bibinfo {year} {2010})}\BibitemShut {NoStop}%
\bibitem [{\citenamefont {Dieck~Kattas}, \citenamefont {Xu},\ and\
  \citenamefont {Small}()}]{Kattas2011b}%
  \BibitemOpen
  \bibfield  {author} {\bibinfo {author} {\bibfnamefont {G.}~\bibnamefont
  {Dieck~Kattas}}, \bibinfo {author} {\bibfnamefont {X.-K.}\ \bibnamefont
  {Xu}}, \ and\ \bibinfo {author} {\bibfnamefont {M.}~\bibnamefont {Small}},\
  }\href@noop {} {\ }\Eprint {http://arxiv.org/abs/1110.1739} {arXiv:1110.1739
  [q-bio.OT]} \BibitemShut {NoStop}%
\bibitem [{\citenamefont {Judd}\ and\ \citenamefont {Mees}(1995)}]{Judd1995}%
  \BibitemOpen
  \bibfield  {author} {\bibinfo {author} {\bibfnamefont {K.}~\bibnamefont
  {Judd}}\ and\ \bibinfo {author} {\bibfnamefont {A.}~\bibnamefont {Mees}},\
  }\href@noop {} {\bibfield  {journal} {\bibinfo  {journal} {Physica D:
  Nonlinear Phenomena}\ }\textbf {\bibinfo {volume} {82}},\ \bibinfo {pages}
  {426} (\bibinfo {year} {1995})}\BibitemShut {NoStop}%
\bibitem [{\citenamefont {Small}\ and\ \citenamefont {Tse}(2002)}]{Small2002}%
  \BibitemOpen
  \bibfield  {author} {\bibinfo {author} {\bibfnamefont {M.}~\bibnamefont
  {Small}}\ and\ \bibinfo {author} {\bibfnamefont {C.~K.}\ \bibnamefont
  {Tse}},\ }\href@noop {} {\bibfield  {journal} {\bibinfo  {journal} {Phys.
  Rev. E}\ }\textbf {\bibinfo {volume} {66}},\ \bibinfo {pages} {066701}
  (\bibinfo {year} {2002})}\BibitemShut {NoStop}%
\bibitem [{\citenamefont {Bongard}\ and\ \citenamefont
  {Lipson}(2007)}]{Bongard2007}%
  \BibitemOpen
  \bibfield  {author} {\bibinfo {author} {\bibfnamefont {J.}~\bibnamefont
  {Bongard}}\ and\ \bibinfo {author} {\bibfnamefont {H.}~\bibnamefont
  {Lipson}},\ }\href@noop {} {\bibfield  {journal} {\bibinfo  {journal}
  {Proceedings of the National Academy of Sciences}\ }\textbf {\bibinfo
  {volume} {104}},\ \bibinfo {pages} {9943} (\bibinfo {year}
  {2007})}\BibitemShut {NoStop}%
\bibitem [{\citenamefont {Gennemark}\ and\ \citenamefont
  {Wedelin}(2007)}]{Gennemark2007}%
  \BibitemOpen
  \bibfield  {author} {\bibinfo {author} {\bibfnamefont {P.}~\bibnamefont
  {Gennemark}}\ and\ \bibinfo {author} {\bibfnamefont {D.}~\bibnamefont
  {Wedelin}},\ }\href@noop {} {\bibfield  {journal} {\bibinfo  {journal}
  {Systems Biology, IET}\ }\textbf {\bibinfo {volume} {1}},\ \bibinfo {pages}
  {120} (\bibinfo {year} {2007})}\BibitemShut {NoStop}%
\bibitem [{\citenamefont {Schmidt}\ and\ \citenamefont
  {Lipson}(2009)}]{Schmidt2009}%
  \BibitemOpen
  \bibfield  {author} {\bibinfo {author} {\bibfnamefont {M.}~\bibnamefont
  {Schmidt}}\ and\ \bibinfo {author} {\bibfnamefont {H.}~\bibnamefont
  {Lipson}},\ }\href@noop {} {\bibfield  {journal} {\bibinfo  {journal}
  {Science}\ }\textbf {\bibinfo {volume} {324}},\ \bibinfo {pages} {81}
  (\bibinfo {year} {2009})}\BibitemShut {NoStop}%
\bibitem [{\citenamefont {Herbert-Read}\ \emph {et~al.}(2011)\citenamefont
  {Herbert-Read}, \citenamefont {Perna}, \citenamefont {Mann}, \citenamefont
  {Schaerf}, \citenamefont {Sumpter},\ and\ \citenamefont
  {Ward}}]{Herbert-Read2011}%
  \BibitemOpen
  \bibfield  {author} {\bibinfo {author} {\bibfnamefont {J.~E.}\ \bibnamefont
  {Herbert-Read}}, \bibinfo {author} {\bibfnamefont {A.}~\bibnamefont {Perna}},
  \bibinfo {author} {\bibfnamefont {R.~P.}\ \bibnamefont {Mann}}, \bibinfo
  {author} {\bibfnamefont {T.~M.}\ \bibnamefont {Schaerf}}, \bibinfo {author}
  {\bibfnamefont {D.~J.~T.}\ \bibnamefont {Sumpter}}, \ and\ \bibinfo {author}
  {\bibfnamefont {A.~J.~W.}\ \bibnamefont {Ward}},\ }\href {\doibase
  10.1073/pnas.1109355108} {\bibfield  {journal} {\bibinfo  {journal}
  {Proceedings of the National Academy of Sciences}\ }\textbf {\bibinfo
  {volume} {108}},\ \bibinfo {pages} {18726} (\bibinfo {year}
  {2011})}\BibitemShut {NoStop}%
\bibitem [{\citenamefont {Cheng}\ \emph {et~al.}(2011)\citenamefont {Cheng},
  \citenamefont {Zhang}, \citenamefont {Chen}, \citenamefont {Zhou},\ and\
  \citenamefont {Valeyev}}]{Cheng2011}%
  \BibitemOpen
  \bibfield  {author} {\bibinfo {author} {\bibfnamefont {Z.}~\bibnamefont
  {Cheng}}, \bibinfo {author} {\bibfnamefont {H.-T.}\ \bibnamefont {Zhang}},
  \bibinfo {author} {\bibfnamefont {M.~Z.~Q.}\ \bibnamefont {Chen}}, \bibinfo
  {author} {\bibfnamefont {T.}~\bibnamefont {Zhou}}, \ and\ \bibinfo {author}
  {\bibfnamefont {N.~V.}\ \bibnamefont {Valeyev}},\ }\href
  {http://dx.doi.org/10.1371%2Fjournal.pone.0022123} {\bibfield  {journal}
  {\bibinfo  {journal} {PLoS ONE}\ }\textbf {\bibinfo {volume} {6}},\ \bibinfo
  {pages} {e22123} (\bibinfo {year} {2011})}\BibitemShut {NoStop}%
\bibitem [{\citenamefont {Ballerini}\ \emph
  {et~al.}(2008{\natexlab{b}})\citenamefont {Ballerini}, \citenamefont
  {Cabibbo}, \citenamefont {Candelier}, \citenamefont {Cavagna}, \citenamefont
  {Cisbani}, \citenamefont {Giardina}, \citenamefont {Orlandi}, \citenamefont
  {Parisi}, \citenamefont {Procaccini}, \citenamefont {Viale},\ and\
  \citenamefont {Zdravkovic}}]{Ballerini2008a}%
  \BibitemOpen
  \bibfield  {author} {\bibinfo {author} {\bibfnamefont {M.}~\bibnamefont
  {Ballerini}}, \bibinfo {author} {\bibfnamefont {N.}~\bibnamefont {Cabibbo}},
  \bibinfo {author} {\bibfnamefont {R.}~\bibnamefont {Candelier}}, \bibinfo
  {author} {\bibfnamefont {A.}~\bibnamefont {Cavagna}}, \bibinfo {author}
  {\bibfnamefont {E.}~\bibnamefont {Cisbani}}, \bibinfo {author} {\bibfnamefont
  {I.}~\bibnamefont {Giardina}}, \bibinfo {author} {\bibfnamefont
  {A.}~\bibnamefont {Orlandi}}, \bibinfo {author} {\bibfnamefont
  {G.}~\bibnamefont {Parisi}}, \bibinfo {author} {\bibfnamefont
  {A.}~\bibnamefont {Procaccini}}, \bibinfo {author} {\bibfnamefont
  {M.}~\bibnamefont {Viale}}, \ and\ \bibinfo {author} {\bibfnamefont
  {V.}~\bibnamefont {Zdravkovic}},\ }\href@noop {} {\bibfield  {journal}
  {\bibinfo  {journal} {Animal Behaviour}\ }\textbf {\bibinfo {volume} {76}},\
  \bibinfo {pages} {201} (\bibinfo {year} {2008}{\natexlab{b}})}\BibitemShut
  {NoStop}%
\end{thebibliography}
%

\end{document}